\documentclass[twoside,9pt]{article}
\usepackage{extsizes}
\usepackage[super,sort&compress,comma]{natbib} 
\usepackage[version=3]{mhchem}
\usepackage[left=1.5cm, right=1.5cm, top=1.785cm, bottom=2.0cm]{geometry}
\usepackage{balance}
\usepackage{mathptmx}
\usepackage{amssymb} 
\usepackage{tcolorbox}
\usepackage{etoolbox}
\usepackage{sectsty}
\usepackage{graphicx} 
\usepackage{lastpage}
\usepackage[format=plain,justification=justified,singlelinecheck=false,font={stretch=1.125,small,sf},labelfont=bf,labelsep=space]{caption}
\usepackage{float}
\usepackage{fancyhdr}
\usepackage{fnpos}
\usepackage[english]{babel}
\addto{\captionsenglish}{%
  
}
\usepackage{array}
\usepackage{droidsans}
\usepackage{charter}
\usepackage[T1]{fontenc}
\usepackage[usenames,dvipsnames]{xcolor}
\usepackage{setspace}
\usepackage[compact]{titlesec}
\usepackage{hyperref}

\usepackage{epstopdf}
\definecolor{cream}{RGB}{222,217,201}

\titleformat{\section}          
  {\bfseries\fontsize{12pt}{14pt}\selectfont} 
  {\thesection}                
  {1em}                        
  {}                           
\begin{document}

\begin{flushleft}
{\fontsize{18pt}{20pt}\selectfont\textbf{Effect of hybrid field coupling in nanostructured surfaces on anisotropic signal detection in nanoscale infrared spectroscopic imaging methods$^\dag$}}
\end{flushleft}
{\fontsize{12pt}{14pt}\selectfont Ayona James,\textsuperscript{a,b} Maryam Ali,\textsuperscript{a,b} Zekai Ye,\textsuperscript{b} Phan Thi Yen Nhi,\textsuperscript{a,b} Sharon Xavi,\textsuperscript{a,c} Mashiat Huq,\textsuperscript{c} Sajib Barua,\textsuperscript{b,c} Meng Luo,\textsuperscript{c} Yisak Tsegazab,\textsuperscript{a,b} Anna Elmanova,\textsuperscript{a,b} Robin Schneider,\textsuperscript{b} 
Olga Ustimenko,\textsuperscript{d}
Sarmiza-Elena Stanca,\textsuperscript{b} Marco Diegel,\textsuperscript{b} Andrea Dellith,\textsuperscript{b} Uwe Hübner,\textsuperscript{b} Christoph Krafft,\textsuperscript{a,b} Jasmin Finkelmeyer,\textsuperscript{b} Maximilian Hupfer,\textsuperscript{a,b} 
Kalina Peneva,\textsuperscript{d,e,f}
Matthias Zeisberger,\textsuperscript{b} Christin David,\textsuperscript{g,h} Martin Presselt,\textsuperscript{a,b}
and Daniela Täuber\textsuperscript{*a,b}}\\\\
Received 00th January 20xx, Accepted 00th January 20xx
DOI: 10.1039/x0xx00000x\vspace{0.5cm}

Anisotropic intensity distributions on nanostructured surfaces and polarization-sensitive spectra have been observed in a number of nanoscale infrared spectroscopic imaging methods, including nano-FTIR [Bakir \textit{et al., Molecules}, 2020, \textbf{25}, 4295], photothermal induced resonance (PTIR) [Waeytens \textit{et al., Analyst}, 2021, \textbf{146}], tapping AFM-IR [Hondl \textit{et al., ACS Meas.\ Sci.\ Au}, 2025, \textbf{5}, 469; Luo \textit{et al., Appl.\ Phys.\ Lett.}, 2022, \textbf{121}, 23330], infrared photoinduced force microscopy (IR-PiFM, PiF-IR) [Anindo \textit{et al., J.\ Phys.\ Chem.\ C}, 2025, \textbf{129}, 4517; Shcherbakov \textit{et al., Rev Methods Primers}, 2025, \textbf{5}, 1; Ali \textit{et al., Anal. Chem.}, 2025, \textbf{97}, 23914] and peak force infrared microscopy (PFIR) [Xie \textit{et al., J.\ Phys.\ Chem.\ C}, 2022, \textbf{126}, 8393; Anindo \textit{et al., J.\ Phys.\ Chem.\ C}, 2025, \textbf{129}, 4517]. A recent work combining modeling and experiment demonstrated that the hybrid field coupling of the IR illumination $E_0$ with a polymer nanosphere and a metallic AFM probe is nearly as strong as the plasmonic coupling in case of a gold nanosphere [Anindo \textit{et al., J.\ Phys.\ Chem.\ C}, 2025, \textbf{129}, 4517]. For p-polarized illumination, this results in enhanced IR absorption on the surface perpendicular to the propagation of $E_0$ which can explain the observed anisotropic intensity distribution. An additional anisotropy may be introduced by aligned surface molecules with oriented vibrational transition moments [Bakir \textit{et al., Molecules}, 2020, \textbf{25}, 4295; Luo \textit{et al., Appl.\ Phys.\ Lett.}, 2022, \textbf{121}, 23330]. PiF-IR is strongly surface sensitive combining an unprecedented spatial resolution < 5 nm with high spectral resolution [Shcherbakov \textit{et al., Rev Methods Primers}, 2025, \textbf{5}, 1; Ali \textit{et al., Anal. Chem.}, 2025, \textbf{97}, 23914], which allows, for example, to visualize nanoscale chemical variation on the surface of bacteria cells affected by antimicrobial interaction [Ali \textit{et al., Anal. Chem.}, 2025, \textbf{97}, 23914]. We compare PiF-IR hyperspectra of aligned perylene Langmuir Blodgett monolayers on nanostructured and planar gold substrates and use quantum chemical calculations of the oriented vibrational oscillators to interpret the observations.
\vspace{1cm}

\section*{Introduction}
The emerging field of nanoscale infrared spectroscopic imaging (NanIR) methods\cite{schwartz_guide_2022, xie_what_2023, shen_scanning_2024} has the potential to revolutionize our understanding of chemical compositions and molecular arrangement in the life and material sciences in a way similar to the breaking insights gained from the emergence of fluorescence microscopy. NanIR methods can bridge the gap between high-resolution structural imaging in electron and atomic force microscopy (AFM) and chemical imaging in conventional far-field infrared (IR) spectroscopy by overcoming the limitations of optical diffraction in far-field IR spectroscopic imaging and provide complementary information to tip enhanced Raman scattering (TERS).\cite{hoppener_tip-enhanced_2024} By exploiting near-field enhanced IR absorption mediated by metal nanostructures, NanIR methods provide access to chemical characterization on the scale of a few nanometers, for example, of antibiotic interaction on the surface of individual bacteria cells.\cite{ali_nanochemical_2025} The highly localized and non-linear absorption is influenced by geometrical parameters in the setup, such as  bottom-up and top-down illumination,\cite{waeytens_probing_2021} the orientation of the linearly polarized IR illumination\cite{waeytens_probing_2021, jahng_quantitative_2022, bakir_orientation_2020} and the geometry of the induced near-field.\cite{pascual_robledo_theoretical_2025} On nanostructured samples, such effects result in anisotropic signal detection, which has been reported on a variety of nanoparticles employing different NanIR methods.\cite{waeytens_probing_2021, shcherbakov_photo-induced_2025, xie_dual-frequency_2022, anindo_photothermal_2025} A better understanding of these effects will result in an improved interpretation of measured signals on nanostructured materials yielding valuable complementary information in a large area of applications.

In general, NanIR methods can be grouped by their approach for signal detection: In surface-enhanced infrared absorption (SEIRA) spectroscopy\cite{griffiths_surface-enhanced_2013, ataka_biochemical_2007} and infrared scattering scanning optical near-field microscopy (IR-SNOM and nano-FTIR)\cite{xu_pushing_2012, huth_resonant_2013, pascual_robledo_theoretical_2025} near-field enhanced absorption is combined with optical far-field detection. In contrast, the so-called AFM-IR methods\cite{xie_what_2023} combine IR absorption enhanced by a sharp metallic tip with mechanical detection via atomic force microscopy (AFM).
Thereby, peak force infrared microscopy (PFIR)\cite{xie_dual-frequency_2022, xie_what_2023} and photothermal induced resonance (PTIR)\cite{dazzi_afmir_2012, waeytens_probing_2021} and its refined methods: resonantly enhanced infrared nanospectroscopy (REINS)\cite{lu_tip-enhanced_2014} and tapping AFM-IR\cite{mathurin_advanced_2020, hondl_method_2025} detect photoinduced thermal expansion upon IR illumination, whereas, in IR photo-induced force microscopy (IR-PiFM or PiF-IR) the photoinduced tip-sample interaction force in the attractive regime is acquired.\cite{jahng_tip-enhanced_2018, sifat_photo-induced_2022, anindo_photothermal_2025, shcherbakov_photo-induced_2025} PiF-IR provides access to chemical characterization of molecular arrangement on surfaces with a spatial resolution of less than 5 nm.\cite{joseph_nanoscale_2024, ali_nanochemical_2025, murdick_photoinduced_2017} This sharp tool can be used to explore optical parameters contributing to field effects and to find tiny contributions from materials (contaminations)\cite{forster_quality_2025} and impacts from approaches in handling the measurements.

Several studies have evaluated hybrid field coupling in NanIR methods for planar layered structures, for example, Jahng \textit{et al.}\ calculated the photoinduced force obtained in PiF-IR for a thin layer of polystyrene on a Si substrate and compared their theoretical results with experimental results.\cite{jahng_nanoscale_2019} Pascual \textit{et al.}\ used electrostatic numerical calculations of infrared near-field spectra in isotropically and anisotropically absorbing dielectric layers on planar \ce{CaF2} and Au substrates, which did not include far-field reflections on the substrate. They found that for very thin layers, the proximity of the Au substrate forces the field lines to align vertically, preventing coupling to in-plane molecular vibrations.\cite{pascual_robledo_theoretical_2025} Xie \textit{et al.}\ applied PFIR and visible peak force microscopy to organo-metal-halide perovsike nanostructures\cite{xie_dual-frequency_2022} and compared their experimental results with near-field modeling. The studied materials show isotropic absorption of the organic compound in the mid-IR and anisotropic visible absorption related to the crystal structure.\cite{tauber_exploring_2016} Anindo \textit{et al.}\ modeled the hybrid photo-induced near-field between a Pt-tip and an isotropically absorbing polymer nanoparticle and compared their results with experimentally obtained results using PiF-IR.\cite{anindo_photothermal_2025} Waeytens \textit{et al.}\ compared bottom-up and top-down PTIR illumination probing amyloid fibril structures with different polarization states of the incoming light field and varying the coating material of the tip both experimentally and with a simplified theoretical approach.\cite{waeytens_probing_2021} Their theoretical model assumes a constant electric incident field and a quasistatic optical response of the involved nanoparticles: a cylindrical fibril and a spherical Au tip. They showed that because of the tip curvature, the components of the electric field change directionality in the vicinity of the tip and can add to the enhancement of absorption in p-polarization while they reduce the field inside the probe for s-polarization,\cite{waeytens_probing_2021} which is in agreement with theoretical considerations of tip-enhanced fields reported by other research groups.\cite{jahng_nanoscale_2019, anindo_photothermal_2025} In amyloid fibrils, proteins organize in a highly oriented $\beta$-sheet structure. Thereby, the transition dipole moment $\mu_\beta$ of the $\beta$-sheet absorption at 1620~cm$^{-1}$ is perpendicular to the fibril’s long axis.\cite{waeytens_probing_2021} For a single fibril with $\mu_\beta$ aligned parallel to s-polarization, Waeytens \textit{et al.}\ found an enhanced signal in the bottom-up illumination for s-polarized light, while the signal in p-polarization could hardly be discriminated from noise.\cite{waeytens_probing_2021} Probing ensembles of arbitrarily oriented amyloid fibrils assembled from three different proteins, they found varying spectral signatures between the three different samples, which can be explained by different percentages of amino acids involved in the formation of $\beta$-sheets in those proteins. For the two proteins that involve a high percentage of amino acids in the formation of $\beta$-sheets, PTIR spectra obtained under bottom-up illumination showed a high intensity in the $\beta$-sheet absorption band, while the intensity of this band was strongly reduced under top-down illumination. In both configurations, Au tips and p-polarization were used. However, for top-down illumination, this results in a higher polarization perpendicular to the sample plane than for bottom-up illumination,\cite{waeytens_probing_2021} while the in-plane components for in-plane polarization are comparable, which might contribute to the observation. 

In this work, we applied high-precision PiF-IR to oriented molecular monolayers of an alkylated perylene-monoimide (PMIS-C8) on nanostructured and planar Au substrates to study the effects of anisotropic absorption in nanostructured surfaces in NanIR imaging experimentally. Asymmetric and amphiphilic perylenes\cite{OPV_Gerase-Amphiphiles-Stable_ApplElectrMat2024} are ideal test systems for this purpose, because their chemical structure is well-defined, with IR-active functional groups firmly anchored to the molecular backbone. Due to their condensed aromatic systems, the number of internal rotational or torsional degrees of freedom is severely limited. The amphiphilicity of the perylenes makes them ideal for producing oriented molecular monolayers using Langmuir-Blodgett (LB) and related methods.\cite{hupfer_supramolecular_2021} We calculated polarization-resolved IR spectra of PMIS-C8 and used them to interpret our experimental results. We further compare the experimentally obtained local PiF intensity variation on nanostructured Au to the field distribution of the photo-induced near-field on a model of the Au nanostructure simulated using a finite element method (FEM) approach.

\section*{Methods}
\subsection*{Au Substrates}
Two types of Au substrates were used: (i) \textit{nanostructured~Au}: 100~nm Au were evaporated onto a 0.6~mm thick \ce{SiO2} chip, resulting in nanostructured Au surfaces consisting of a dense population of $\approx 15$~nm high and $\approx 50$~nm wide hills and (ii) \textit{planar Au}: 200~nm Au were evaporated onto a \ce{Si} wafer. Subsequently, a \ce{SiO2} chip was placed onto the Au layer. When the chip was carefully removed, the Au layer adhered to the \ce{SiO2} exposing the smooth surface that had been formed at the interface with the \ce{Si} wafer.

\subsection*{Perylene}
The \ce{N-Octylmonoimide} of \ce{9,10-bis([1,2]dithiole)-1,6,7,12-tetrachloroperylene} (PMIS-C8) was sythesized as described perviously.\cite{abul-futouh_toward_2018, hupfer_supramolecular_2021}
For comparative infrared spectroscopy, PMIS-C8 was dissolved in \ce{CCl_4} with a concentration of (0.339~µmol~mL$^{-1}$).

\subsection*{Langmuir-Blodgett (LB) Films}
LB films were prepared following established procedures\cite{OPV-SupraMol_Das-ELUMOVOC_AdvEMat2018,SupraM-Thiaz_Hupfer-ThiazAlkAryl_Langmuir2019,SupraM-Thiaz_Hupfer-ThiazCentralLinker_ACSApplMatInterf_2017} using a PMIS-C8 dichloromethane solution (0.1~µmol~mL$^{-1}$) and a target surface pressure of 20~mN~m$^{-1}$. The spread volume was 700~$\mu$L and 900~$\mu$L for nanostructured and planar Au substrates, respectively.

Film deposition for the two final samples on nanostructured Au was carried out using a KSV NIMA Langmuir and Langmuir Blodgett trough in the education lab at the Institute of Physical Chemistry. For all other samples, a KSV 5000 Langmuir Blodgett trough available in the Presselt lab was used. In the KSV 5000, two substrates were mounted simultaneously by attaching each to a clip in the sample holder. The barrier compression speed was set at 10~mm~min$^{-1}$, corresponding to a compression rate of approximately 5~mN~m$^{-1}$ min$^{-1}$. Film formation was monitored using Brewster Angle Microscopy (BAM). In the smaller KSV NIMA, only one sample was mounted and the barrier compression speed was set at 3.5~mm~min$^{-1}$, corresponding to the same compression rate of approximately 5~mN~m$^{-1}$ min$^{-1}$. In both instruments, the substrates were pulled out at a speed of 2~mm~min$^{-1}$ during deposition and the transfer of the monolayer to the Au substrates was carried out at a surface pressure of 20~mN~m$^{-1}$.

\subsection*{Brewster Angle Microscopy (BAM)}
BAM measurements were carried out using a KSV NIMA MicroBAM system (Biolin Scientific, Finland) equipped with a 659~nm laser illumination with 50~mW maximum laser power at the aperture, a microscope objective lens and a CCD detector. BAM images were recorded every 20~min on a 3580~µm wide area in situ during monolayer formation at the air–water interface. Exemplary BAM images of a PMIS-C8 film formed at the water surface are shown in the supporting information in Figure S1.

\subsection*{Atomic Force Microscopy (AFM)}
AFM height and phase images were acquired in air in tapping mode ($f_t=197$~kHz or $f_t=307$~kHz) on a Bruker-AFM Dimension Edge\texttrademark~(BrukerAXS, Karlsruhe, Germany). 20~$\mu$m wide overviews and 5~$\mu$m wide higher resolution scans were acquired with scan resolutions of 39~nm/pixel and 9.8~nm/pixel, respectively. AFM data were processed using the following tools in Gwyddion\cite{necas_gwyddion_2012}: (i) Level data by mean plane subtraction, (ii) Align rows using various methods (applying a 5th degree polynomial) and (iii) Correct horizontal scars. Height images were additionally processed using the tool (iv) Level data to make facets point upward. AFM height and phase images of pristine Au substrates and PMIS-C8 monolayer films are presented in the supporting information in Figures~S2 and S3, respectively.

\subsection*{Mid-IR Photo-induced Force Microscopy (PiF-IR)}
PiF-IR was conducted on a commercial VistaScope\texttrademark~(Molecular Vista, US) equipped with a pulsed quantum cascade laser (MirCat\texttrademark, Daylight Solutions, US), which was tuned in the spectral range of 800 - 1800~cm$^{-1}$ and operated in side band mode.\cite{sifat_photo-induced_2022,anindo_photothermal_2025} For this, the pulse frequency $f_m$ of the MirCat\texttrademark~was tuned to match $f_m=f_2-f_1$, where $f_1$ and $f_2$ are the first and second mechanical resonance frequencies of the \ce{PtIr}-coated AFM cantilever (Point Probe Plus, Nanosensors, CH). The photo-induced force (PiF) was acquired in the side band mode at $f_1$, while a second login amplifier was used to record the topography at the driving frequency $f_2$. Due to the smaller amplitude of the oscillation at $f_2$, the topography contrast is more sensitive to instrument noise than the PIF contrast acquired at $f_1$. In several positions, height differences caused by tilt in sample mounting exceeded the small height difference of 10-30~nm in our samples. These topography images were processed using line averages in the fast scanning (horizontal) direction. PiF contrasts were not corrected.

\begin{figure}[ht]
\centering
  \includegraphics[height=10 cm]{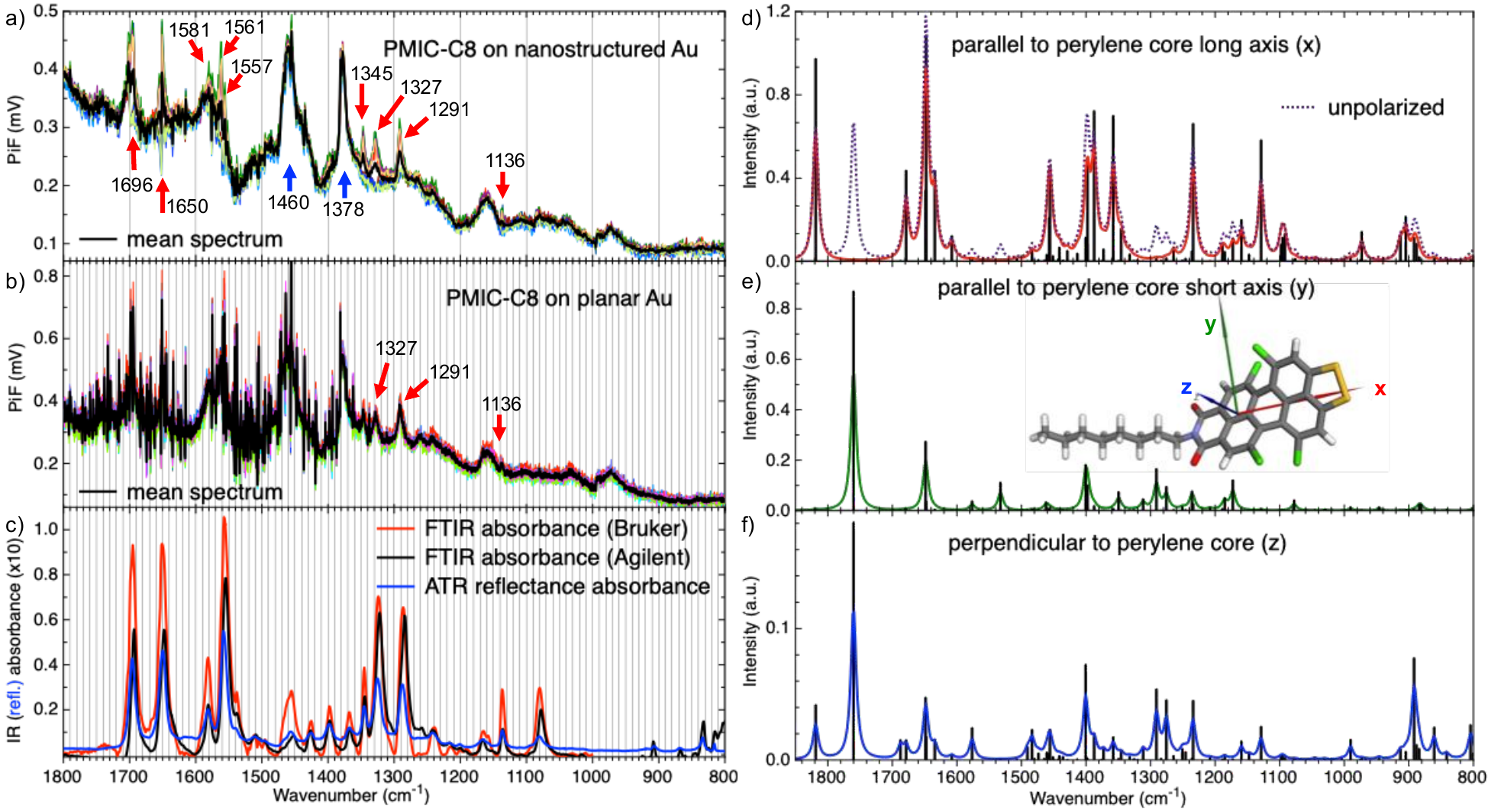}
  \caption{{\bf Experimental and computed infrared spectra of PMIS-C8.} a) and b) PiF spectra of PMIS-C8 monolayer on nanostructured and on planar Au, respectively; red and blue arrows in a) mark bands showing intensity variations and no variations, respectively; c) FTIR and ATR spectra; d-e) computed stick and line broadened IR spectra: d) unpolarized line broadened spectrum (dotted line) and component spectra parallel to long axis (x) of perylene core (solid line, and stick spectrum), e) component spectra parallel to short axis (y) of perylene core and f) component spectra perpendicular (z) to perylene core.}
  \label{fig1}
\end{figure}
The illumination power was homogenized over the spectral range by clipping the illumination intensity to manually selected percentages of peak values in previously obtained power spectra. Depending on the selected value, this resulted in an illumination power in the range of 20-40~$\mu$W with a focal spot diameter of $\approx2\lambda$. For this procedure, power spectra are acquired in direct mode at $f_1$ and $f_2$ in air. A "side-band" power spectrum is then calculated from the obtained oscillation amplitudes at both frequencies.\cite{jahng_quantitative_2022} The quality of the homogenization strongly depends on the stability of the IR beam alignment to the tip-sample region. To recover the alignment after mounting new samples, we first validated the alignment in air and then acquired new power spectra before starting data acquisition. The instrument was purged with \ce{N2} to reduce the contributions of water vapor absorption to the acquired spectra. Nevertheless, slight variations in humidity between the acquisition of the power spectra and subsequent data acquisition may result in residual contributions of water vapor lines to the acquired spectra. Since we installed the new powerful MirCat light source in the instrument, we frequently have observed rather strong baselines that rise from lower to higher frequencies. Two examples are seen in the PiF-IR spectra in Figure~\ref{fig1}a,b. The cause of this rise is not yet fully known and there is a nonlinear dependence of the PiF on illumination power. Thus, we decided not to apply any math operations to remove this rise.

PiF-IR single point spectra are usually acquired on selected positions in a previously scanned sample area. Although the high-precision closed loop scanner in the instrument causes only small drifts of a few nanometers between succeeding single frequency illumination scans of the same sample area, consecutive moves to selected positions for point spectra will result in an accumulated drift on the scale of the structural variation in the nanostructured Au substrate. In hyperspectral PiF-IR scans, much higher positional accuracy is provided, enabling the comparison of spectral bands at particular sample positions. Due to the much longer acquisition time ($\approx 20$~h), the instrument drift accumulates over the hyperscan, resulting in scan areas elongated in the slow scan direction compared to much faster single frequency scans (20 - 120~min). 

\textbf{PiF-IR data analysis} of single frequency scans, hyperspectral scans, and spectra was conducted using the instrument software SurfaceWorks\texttrademark~(Molecular Vista, US). At interesting vibrational frequencies $\nu$, spectral channels with bandwidth $\nu\pm2$~cm$^{-1}$ were extracted from hyperspectral scans.
Hierarchical Cluster Analysis (HCA) of PiF-IR hyperspectra was performed using our home-written software hyPIRana.\cite{ali_hypirana_2026} Three HCA clusters were obtained from the dendrograms. For each cluster, mean spectra (HC1, HC2 and HC3) were calculated. We applied a non-negative factorization to obtain cluster maps to study the relative contribution of each cluster in each spectrum (S) of the hyperspectral data set: $\mathrm{S} = \alpha_1\mathrm{HC1} + \alpha_2\mathrm{HC2} + \alpha_3\mathrm{HC3}$. The resulting factor maps show relative contributions of the three components in each pixel.
PiF-IR is highly sensitive to contaminants.\cite{forster_quality_2025}  Polydimethylsilane (PDMS), a highly volatile polymer, was frequent in our lab. We used its characteristic absorption at 1262~cm$^{-1}$ to check for sample contamination by PDMS.

\subsection*{Conventional infrared (IR) spectroscopy}
\textbf{Attenuated total reflection (ATR) spectroscopy} was conducted on a commercial instrument (model 670, Agilent, US) at a spectral resolution of 2~cm$^{-1}$. 30~$\mu$l PMIS-C8 solution was dripped onto a \ce{ZnSe} crystal. ATR spectra were acquired after evaporation of \ce{CCl_4}. The spectra were calibrated from a background taken before droplet disposal and corrected for multiple reflection using the instrument software (Resolution Pro v.5.3).

\textbf{Fourier transform infrared (FTIR) spectra} were acquired from a dried droplet of 10~$\mu$l PMIS-C8 solution dripped on a \ce{CaF2} window (Crystec, Berlin) at a spectral resolution of 2~cm$^{-1}$, 32 spectra per probe in the spectral range of 600 – 4000~cm$^{-1}$ in  transmission mode using two different commercial instruments:
(i) Hyperion microscope (Bruker, US) equipped with a Cassegrain 15×/NA 0.4 objective connected to a Bruker Vertex 80v spectrometer. The samples were illuminated with a silicon carbide (\ce{SiC}) infrared light source and the spectra were collected with a standard FTIR detector with Mercury Cadmium Telluride (\ce{MCT}) diode ($D^*:>2$ × 1010~cmHz$^{1/2}$W$^{-1}$) liquid nitrogen cooled. The measurements were performed at the source aperture of 4~mm, \ce{KBr} beam splitter, collection mirror velocity 20~kHz. The spectra were recorded on a square sample spot measuring 400~$\mu$m x 400~$\mu$m. Reference for transmission mode: Air. OriginPro\texttrademark~was used for baseline correction.
(ii) Agilent model 670 (Agilent, US) equipped with a Cassegrain 15×/NA 0.4 objective. The spectrometer chamber was enclosed in a homebuilt box and purged by dry air to reduce the spectral contributions of water vapor. Instrument software (Resolution Pro v.5.3) was used to calibrate the FTIR spectra from a background taken on a clean substrate. Previously acquired water vapor spectra were used to remove residual water lines from the spectra. 

\subsection*{Computation of IR spectra}
\subsubsection*{Quantum chemical calculation}
Quantum chemical structure optimizations along with the calculations of vibrational spectra were carried out using density functional theory (DFT) as implemented in the GPU-accelerated program TeraChem.\cite{ufimtsev_quantum_2009, isborn_excited-state_2011, titov_generating_2013, song_automated_2016, beenken_origin_2015, song_efficient_2015, presselt_prediction_2009} To account for non-covalent interactions, Grimme’s D3 dispersion correction was employed for all calculations.\cite{grimme_consistent_2010, das_macroscopic_2020} All geometries were optimized at the CAM-B3LYP\cite{yanai_new_2004} level of theory with the valence double-$\zeta$ basis set Ahlrichs-pVDZ, including polarization functions on all atoms as introduced by Ahlrichs and co-workers.\cite{schafer_fully_1992} The optimized structures were confirmed as true minima through frequency calculations.

\subsubsection*{Post-processing and vibrational analysis}
The calculated vibration data were further processed using a custom script to determine IR intensities and generate theoretical spectra.

\textbf{Data Extraction and Coordinate System Definition}: The script parsed the final optimized geometries, Atomic Polar Tensors (APTs), and mass-weighted normal modes from the TeraChem output files. To establish a consistent frame of reference, the molecule was oriented into its standard orientation, placing the center of mass at the origin and aligning the Cartesian axes with the principal axes of inertia. This procedure aligns the $x$-axis with the long axis of the planar aromatic core, the $y$-axis orthogonal to the molecular plane, and the $z$-axis within the plane, perpendicular to the $x$-axis. This well-defined coordinate system served as the foundation for all subsequent vector analysis.

\textbf{Calculation of Cartesian Displacements and IR Intensities}: From the calculation output, the Atomic Polar Tensor (APT) and mass-weighted normal modes were extracted. The mass-weighted modes were subsequently converted to Cartesian displacement vectors (I) to present the true physical motion of the atoms. The transition dipole moment vector ($\mu$) for each mode, also known as IR-transition vector, was calculated by the matrix product of the APT and the corresponding displacement vector ($\mu= APT\times I$). The total IR intensity was then derived from the squared magnitude of this vector ($\mu^2 = \mu_x^2 + \mu_y^2 + \mu_z^2$), and its individual components were also resolved to understand the directional character of each absorption.

\textbf{Generation of Theoretical IR Spectra}: The computed vibrational frequencies and corresponding IR intensities were used to construct discrete IR stick spectra. To simulate experimental absorption profiles, each transition was convoluted with a Lorentzian line-shape function using a full width at half maximum (FWHM) of 10.0~cm$^{-1}$. This yielded a broadened theoretical IR spectrum. To further examine the directional character of absorption, the squared Cartesian components of the transition dipole moment ($\mu_x^2$, $\mu_y^2$, $\mu_z^2$) were individually convoluted using the same broadening function to generate component-resolved spectra for the $x$, $y$, and $z$ directions.

\textbf{Visualization of Vibrational Modes}: To facilitate mode assignment, interactive 3D visualizations were created using the py3Dmol library,\cite{seshadri_3dmoljs_2020} allowing inspection of static atomic displacement vectors and dynamic animations of vibrational motions.

\subsection*{Modeling hybrid field coupling}
In this work, the optical properties of nanostructures on substrates are simulated using the finite element method (FEM) implemented in COMSOL Multiphysics 6.3 using the wave optics module in the frequency domain. The refractive index of gold (Au) used for the modeling is taken from Olmon \textit{et al.}\cite{olmon_optical_2012} covering wavelengths from 3 to 25~$\mu$m well into the infrared. An Au nanostructure is placed on a 100~nm thick planar Au layer on top of a 100~nm thick \ce{SiO2}~\cite{franta2016optical} layer. Floquet periodic boundary conditions were applied to all four surrounding surfaces to simulate a periodic array of nanosized unit cells. The excitation light wave is p-polarized (TM) and incident at an angle of $60^\circ$.

The Au nanostructure has the shape of a $\cos^2(\pi x/R)$ function with radius of 25~nm and a height of 15~nm and the width of the unit cell being $2R=50$~nm. This 2D parametric curve is then revolved around the $z$-axis to obtain a 3D nanostructure. A very fine mesh with elements of an average size of $2.5$~nm was used on the surface of the curved structure. Further details on the simulations can be found in the supporting information.

\section*{Results}
\subsection*{Intensity variations of characteristic PMIS-C8 bands in PiF-IR spectra}

Single-point PiF-IR spectra acquired at varied positions on a PMIC-C8 monolayer on nanostructured Au show intensity variations in several absorption bands, which are marked by red arrows in Figure~\ref{fig1}a. These bands are related to vibrations in the imide group (antisymmetric and symmetric \ce{C=O} stretch at 1696 and 1650~cm$^{-1}$, respectively)\cite{del_cano_molecular_2004} and the perylene core (\ce{C=C} in-plane stretch at 1581~cm$^{-1}$,\cite{del_cano_molecular_2004} \ce{C=C} out-of-plane stretch at 1561 \& 1557~cm$^{-1}$,\cite{lapinski_vibrational_2006} \ce{C=C} and \ce{CN} in-plane stretch at 1345~cm$^{-1}$,\cite{del_cano_molecular_2004} and \ce{CH} bend at 1327 \& 1291~cm$^{-1}$\cite{lapinski_vibrational_2006}). In contrast, the two strong bands appearing at 1460 and 1378~cm$^{-1}$ (blue arrows in Figure~\ref{fig1}a) show no such variations above noise level. These two bands are related to \ce{CH3} and \ce{CH2} bending\cite{coates_interpretation_2006} in the alkyl chain. 
All these bands also appear in PiF-IR single-point spectra acquired at several positions on a PMIS-C8 monolayer on planar Au; see Figure~\ref{fig1}b. However, the spectra of this sample show strong contributions by water vapor in the spectral range between 1800 and 1350~cm$^{-1}$. 
The PiF signal obtained from the monolayer on planar Au was considerably weaker than that obtained from PMIS-C8 on nanostructured Au. We therefore had increased the illumination power from $\approx20$~mW to $\approx40$~mW, which resulted in comparable PiF intensities, in agreement with the ratio of average field intensities of photoinduced fields on planar and nanostructured Au obtained from modeling; see Table~S1 in the supporting information. Unfortunately, the increased illumination power also increased the sensitivity of the PiF signal to contributions from incomplete compensation of water vapor, resulting in noisy spectra in this spectral region.
Nevertheless, the bands at 1327, 1291 and 1136~cm$^{-1}$ related to perylene core vibrations are visible outside this noisy region; see red arrows in Figure~\ref{fig1}b. In contrast to the PiF spectra obtained from PMIS-C8 on nanostructured Au (Figure~\ref{fig1}a), all PiF spectra in this data set show similar intensities in these three bands, reporting a rather homogeneous orientation of the perylene core in the investigated positions of the PMIS-C8 monolayer on the planar Au substrate. We will discuss molecular orientation in more detail after a comparison of PiF-IR spectra with conventional IR spectra.

\begin{table*}[h!]
\small
  \caption{Vibrational bands of PMIC-C8 and their assignment to vibrations of imide (I), perylene core (P) and alkyl chain (A).}
  \label{Table1}
  \begin{tabular*}{0.9\textwidth}{@{\extracolsep{\fill}}cccccc}
    \hline\\
    PiF-IR & ATR & FTIR & computed & band assignment & $\mu\|$ to axis \\
    ($\pm1$~cm$^{-1}$) & ($\pm1$~cm$^{-1}$) & ($\pm1$~cm$^{-1}$) & (cm$^{-1}$) &&\\
    \\
    \hline\hline\\
   1696 & 1696 & 1693 & 1819 & \ce{C=O} (I) antisym. stretch\cite{del_cano_molecular_2004} & \textbf{x}, z \\
    \\
    \hline\\
   1650 & 1650 & 1648 & 1760 & \ce{C=O} (I) sym. stretch\cite{del_cano_molecular_2004} & \textbf{y}, z \\
    \\
    \hline\\
     1581 & 1581 & 1581 & 1679 & \ce{C=C} (P) in-plane stretch\cite{del_cano_molecular_2004} & \textbf{x}, z \\
     &&& 1648 && \\
     &&& 1634 && \\
    \\
    \hline\\
     1561& 1558 & 1555 & 1608 & \ce{C=C} (P) out-of-plane stretch\cite{lapinski_vibrational_2006} & \textbf{y}, z \\
     1557&&& 1533 && \\
    \\
    \hline\\
     1460& 1457 & 1452 & 1456 & \ce{CH3} and \ce{CH2} (A) bending\cite{coates_interpretation_2006} & \textbf{x}, z, y \\
     & 1425 & 1426 & 1401 && \\
     1378 & 1398 & 1397 & 1398 && \\
     1366 & 1367 & 1367 & 1388 && \\
    \\
    \hline\\
     1345 & 1345 & 1344 & 1358 & \ce{C=C}, \ce{CN} (P) in-plane stretch\cite{del_cano_molecular_2004} & \textbf{x} \\
     \\
    \hline\\
     1327 & 1325 & 1324 & 1346 & \ce{CH} (P) bend \& & \textbf{x} \\
     &&& 1332 & \ce{CC} def. (A)\cite{lapinski_vibrational_2006}& \textbf{x}\\
     \\
    \hline\\
     1291 & 1288 & 1284 & 1291 & \ce{CH3} and \ce{CH2} (A): H-C-H twist\cite{lapinski_vibrational_2006} \& & \textbf{y}, z, x \\
     &&& 1276 & \ce{CH} (P) bend & \textbf{y}, z \\
     \\
    \hline\\
     1136 & 1136 & 1136 & 1129 & \ce{CH} (P) and \ce{CH2} (A) & \textbf{x}, z \\
    \\
    \hline
  \end{tabular*}
\end{table*}

The vibrational bands of PMIS-C8 discussed above are also visible in conventional ATR and FTIR spectra; see Figure~\ref{fig1}c. Table~\ref{Table1} provides a comparison of band positions of the major PMIS-C8 bands for the three experimental methods, including also calculated spectra. In general, the band positions agree quite nicely for the three experimental methods. The band positions of the calculated spectra show considerable deviations from the experimental ones, in particular for the antisymmetric and symmetric carbonyl vibrations of the imide moiety, which appear in the experimental spectra at $\approx1696$ and $\approx1650$~cm$^{-1}$, respectively. Calculations were conducted for a single PMIS-C8 molecule assuming a polar environment, which does not fully reproduce experimental conditions. In particular, PMIS-C8 is known to form dimers in monolayer films on silver (\ce{Ag}) substrates.\cite{hupfer_embedding_2021}
As can be seen in Figure~\ref{fig1}c, the band positions slightly differ between ATR and FTIR spectra acquired on the same instrument (Agilent spectrometer) and also between the FTIR spectra acquired on two different instruments. In addition, relative band intensities vary considerably between spectra acquired using different methods or instruments. Such variations are expected from different illumination geometries and baseline corrections, which can also explain the slight variations between band positions in conventional IR spectra and PiF-IR spectra. The cause of the rather strong baselines rising from lower to higher frequencies in PiF-IR spectra (Figure~\ref{fig1}a,b) is not fully understood, for details see the methods section.

A remarkable difference between PiF-IR spectra of monolayer PMIS-C8 films and conventional IR spectra acquired on PMIS-C8 droplets on \ce{CaF2} after solvent evaporation are the two strong and broad bands at 1460 and 1378~cm$^{-1}$ in the PiF-IR spectra (blue arrows in Figure~\ref{fig1}a). Using quantum chemical calculations, we found several \ce{CH3} and \ce{CH2} bending vibrations of the alkyl chain in this spectral region; see table~\ref{Table1}
in agreement with literature reports.\cite{coates_interpretation_2006} In the conventional IR spectra, these bands are much weaker compared to the other bands, and several sub-bands are resolved (Figure~\ref{fig1}c), which seem to appear as shoulders in the broad bands in the PiF-IR spectra (Figure~\ref{fig1}a,b). Two effects contribute to this observation: (i) PiF-IR is surface sensitive and PMIS-C8 is expected to bind to the Au substrate through its thiol groups, while the alkyl chain forms the interface with air\cite{hupfer_supramolecular_2021} facing the AFM tip, and (ii) PiF-IR is sensitive to field enhancement and the direction of force detection perpendicular to the sample stage\cite{jahng_nanoscale_2019, sifat_photo-induced_2022, anindo_photothermal_2025} and the alkyl chains are flexible, performing \ce{CH3} and \ce{CH2} bend vibrations more freely in space in contrast to vibrations of the perylene core.\cite{hupfer_supramolecular_2021} This mobility of the alkyl chain was not included in the calculation of polarization-resolved IR spectra presented in Figure~\ref{fig1}d-f, which show a stronger intensity of these bands in the direction parallel to the perylene core long axis ($x$-component). IR spectra were calculated using a lower energy confirmation of the molecular structure of PMIS-C8\cite{hupfer_supramolecular_2021} showing a strict and straight orientation of the alkyl chain (see inset in Figure~\ref{fig1}e). This is not necessarily the case on the surface of the PMIS-C8 monolayer films at room temperature, resulting in a decrease in the orientational anisotropy in these bands compared to the results of the calculation. The perylene core is less flexible in re-orientation within monolayer films.\cite{hupfer_supramolecular_2021} The high spatial resolution of PiF-IR can resolve local intensity variations in bands related to perylene core vibrations (see Figure~\ref{fig1}a). The scan areas for PiF-IR spectra shown in Figure~\ref{fig1}a,b are presented in the supporting information together with arrows marking the positions of spectral acquisition and two further investigated sample positions. We will use hyperspectral PiF-IR scans for studying correlations of spectral bands and sample positions because of their higher positional accuracy which is detailed in the methods section. 

\subsection*{PiF-IR probes molecular orientation in PMIS-C8 monolayers on Au substrates}
\begin{figure}[th]
\centering
  \includegraphics[height=13 cm]{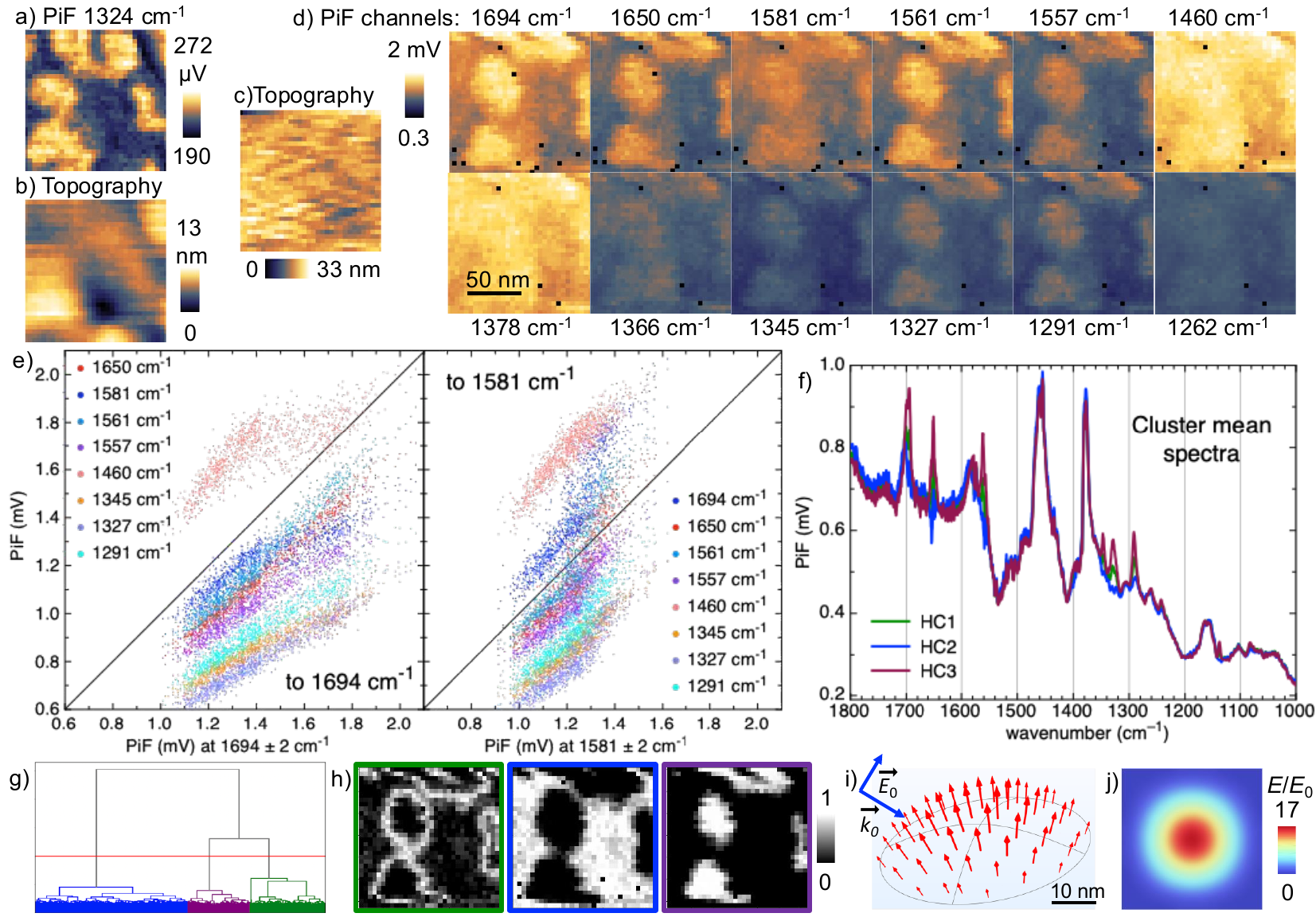}
  \caption{{\bf Local molecular orientation of PMIS-C8 on nanostructured Au substrate visualized using PiF-IR hyperspectral bands.} a) PiF and b) simultaneously acquired topography in single frequency scan at $\nu=$~1324~cm$^{-1}$, c) topography acquired with PiF-IR hyperscan in the same area, d) PiF contrast of selected bands in the hyperscan, e) PiF intensity correlations of characteristic PMIS-C8 bands of the hyperscan to the bands at 1694~cm$^{-1}$ (left) and 1581~cm$^{-1}$ (right); f-h) HCA analysis of hyperspectrum: f) cluster mean spectra, g) dendrogram and h) factor maps in matching colors, i,j) modeled photo-induced $E$-field on Au nanostructure: i) orientation and j) normalized field strength in sample plane.}
  \label{fig2}
\end{figure}
\textbf{PIF contrasts acquired on PMIS-C8 monolayer films on nanostructured Au} at illumination frequencies related to vibrations of the perylene core show local intensity modulations, an example is seen in Figure~\ref{fig2}a. The intensity variations roughly match the sample topography in this position (Figure~\ref{fig2}b), which shows the typical surface structure of the evaporated Au layer of the underlying nanostructured Au substrate. We acquired a PiF-IR hyperscan in this area to study correlations between band intensities. As detailed in the methods section, the simultaneously acquired topography image (Figure~\ref{fig2}c) is rather noisy due to long acquisition, while the PIF channels (Figure~\ref{fig2}d) are less sensitive to instrument noise. The structure seen in the PiF channels, which are related to vibrations of the imide group (1694 and 1650~cm$^{-1}$) and the perylene core (1561, 1557, 1345, 1327 and 1291~cm$^{-1}$), follows the PiF contrast seen in the single frequency scan (Figure~\ref{fig2}a) of this sample position. The slight distortion of the structure between the two scans results from an accumulation of positional drift in the instrument over the long acquisition time of the hyperscan. In contrast, PiF channels connected with a substantial amount of \ce{CH2} and \ce{CH3} bend vibrations show less contrast, in particular those acquired at 1460 and 1378~cm$^{-1}$. The low intensity and low contrast in the channel at 1262~cm$^{-1}$ confirms the absence of the frequent contaminant PDMS in this sample position.

A pixel-wise correlation of PiF-intensities in characteristic bands (Figure~\ref{fig2}e) provides further details of the correlations between the PMIS-C8 absorption bands. Below a certain threshold, ($\approx 1.4$~mV for the channel at 1694~cm$^{-1}$ and $\approx 1.2$~mV for the channel at 1581~cm$^{-1}$ in Figure~\ref{fig2}e left and right, respectively) all PiF-intensities show a parallel rise. Above this threshold, PiF intensities stay almost constant in the two channels at 1581 and 1460~cm$^{-1}$, while they continue to increase in parallel in the channels at 1694, 1650, 1561, 1557, 1345, 1327, and 1291~cm$^{-1}$.
This correlation of vibrational bands with $\mu$ oriented along the molecular $x$ and $y$-axes (see Table~\ref{Table1}) shows that vibrations parallel to the perylene core's long axis cannot be discriminated from those parallel to its short axis. Seemingly, in positions related to PiF intensities above the threshold, the molecules orient mostly with its core perpendicular to the substrate, but without a definite orientation of their long axes. This interpretation is consistent with the formation of stacked PMIS-C8 dimers in monolayer LB films on Ag substrates.\cite{hupfer_supramolecular_2021} Due to the twist in the perlene core caused by the large chlorine atoms in the bay positions, the stacks only form between half of the core.\cite{hupfer_supramolecular_2021}  As a consequence, there is a tilt angle between the molecular long axes and a turn between the imide group orientations, which can explain the correlation between intensities of molecular vibrations oriented parallel to the long and short axes of the perylene core.

Further insights on local molecular orientation can be derived from an HCA analysis of this hyperspectral scan (Figure~\ref{fig2}f-h). The cluster mean spectra (HC1, HC2 and HC3 in Figure~\ref{fig2}f) calculated for the selected clusters in the dendrogram (Figure~\ref{fig2}g) differ in the intensities of bands that are sensitive to molecular orientation (imide group vibrations at 1694 and 1650~cm$^{-1}$ and perylene core vibrations at 1561, 1557, 1345, 1327 and 1291~cm$^{-1}$). Factor maps were derived from a linear combination of HC1, HC2 and HC3 in each pixel (for details, see the methods section). They show a correlation of the relative intensities of the three components (presented in matching colors in Figure~\ref{fig2}h) with the sample topography: HC3 relates high intensities in the bands sensitive to molecular orientation (Figure~\ref{fig2}f) with elevated regions in the sample topography (compare Figures~\ref{fig2}b,c and h taking also into account the distortion of the scanned area during the hyperscan acquisition). In HC2 the orientation sensitive bands show low intensities, suggesting an orientation of the perylene core mostly parallel to the sample surface. A high factor of HC2 correlates mainly with lower regions in the sample topography, which also show lower PiF intensities in hyperspectral channels related to imide and perylene core vibrations in Figure~\ref{fig2}d. A high factor of HC1 (Figure~\ref{fig2}g) marks the contour of elevated areas in the sample topography. HC1 shows intermediate intensities in the orientation sensitive spectral bands compared to the other two mean cluster spectra (Figure~\ref{fig2}f). Thus, its distribution can be assigned to regions in the sample, where the perylene cores turn from an orientation perpendicular to the sample plane to an in-plane orientation. This interpretation is further supported by results from modeling the photo-induced field on a model nanostructure: The field vectors are oriented normal to the surface of the nanostructure while the field intensity shows a decrease from its top region to its edge; see Figure~\ref{fig2}i and j, respectively. For details of the optical modeling, we refer to the methods section and the supporting information. The hyperscan covers only a small sample area containing a few "hills" and "valleys" in the sample topography. We will investigate molecular orientation and field effects in more detail using single frequency scans covering a larger scan area in the next section after presenting results of a comparable hyperscan obtained from a PMIS-C8 monolayer on planar Au. 

\begin{figure}[htb]
\centering
  \includegraphics[height=13 cm]{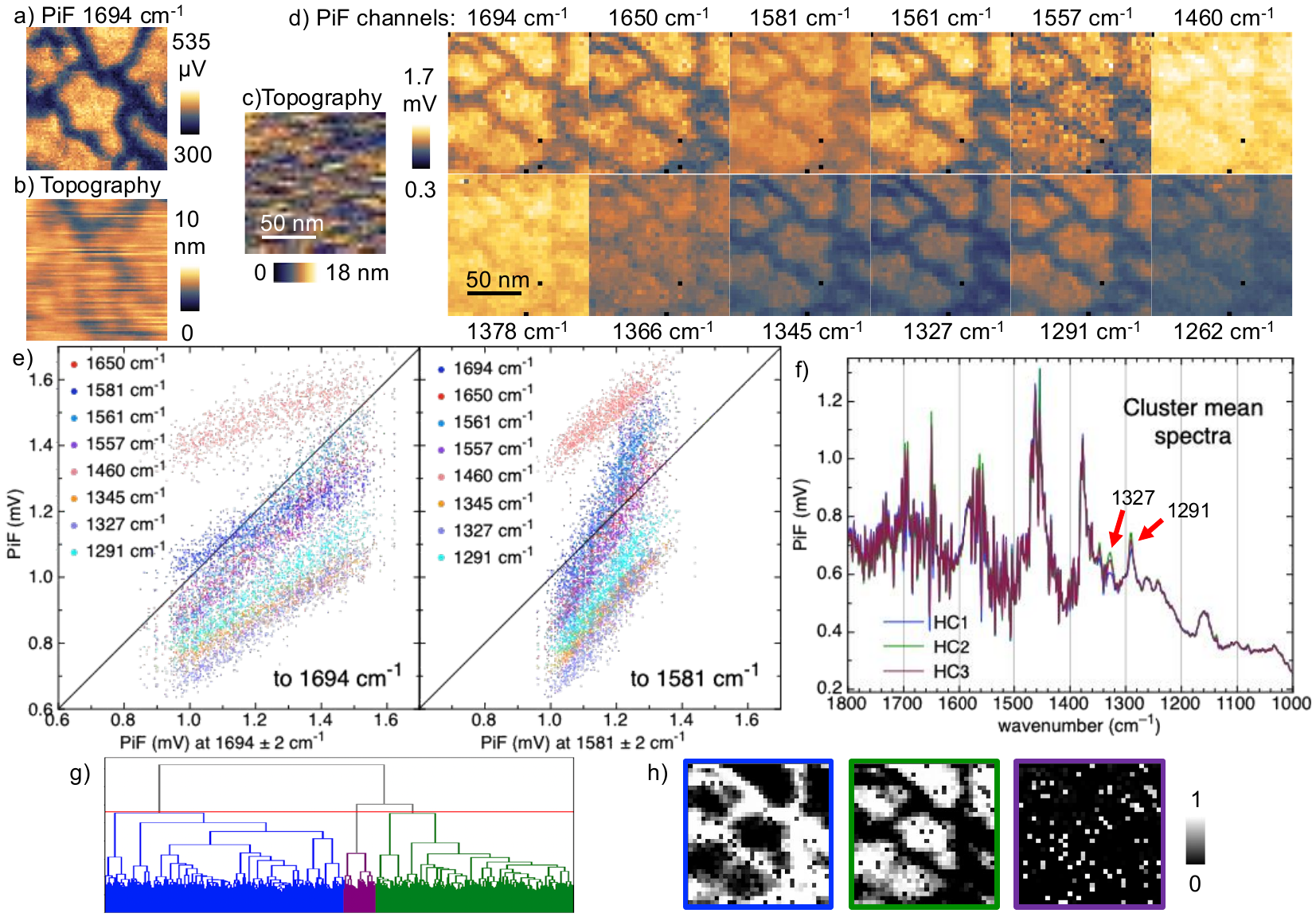}
  \caption{{\bf Local molecular orientation of PMIS-C8 on planar Au substrate visualized using PiF-IR hyperspectral bands.} Selected areas of a) PiF and b) simultaneously acquired topography in single frequency scan at $\nu=$~1694~cm$^{-1}$ shown in Figure S5, c) topography acquired with PiF-IR hyperscan in the same area, d) PiF contrast of selected bands in the hyperscan, e) PiF intensity correlations of characteristic PMIS-C8 bands of the hyperscan to the bands at 1694~cm$^{-1}$ (left) and 1581~cm$^{-1}$ (right); f-h) HCA analysis of hyperspectrum: f) cluster mean spectra with characteristic bands marked by red arrows, g) dendrogram and h) factor maps.}
  \label{fig3}
\end{figure}

\textbf{Planar Au substrates} contain planar areas with only few nm height variation, which are surrounded by small steps and small particles, as can be seen in the AFM investigation of a 5~$\mu$m wide area presented in Figure~S2a in the supporting information. PMIS-C8 monolayers form a pattern of up to 100 nm wide planar flakes separated by small valleys and dips (see Figure S3a and b). PiF contrasts at $\nu=1324$~cm$^{-1}$ and $\nu=1694$~cm$^{-1}$ show an increased intensity in the planar areas compared to the valleys; see overview scans in Figure S4. We selected a 130 nm wide area for a PiF-IR hyperscan, which covered several planar areas and valleys from these overview scans; matching cuts of PiF contrast and topography from the scan at 1694~cm$^{-1}$ are presented in Figure~\ref{fig3}a and b, respectively. The topography image (Figure~\ref{fig3}c) acquired simultaneously with the hyperscan is too noisy to discriminate its structure. Thus, we will again rely on the topography (Figure~\ref{fig3}b) of the previously acquired single frequency scan to relate PiF contrasts in selected channels of the hyperscan (Figure~\ref{fig3}d) with sample topography. Similar to PMIS-C8 on nanostructured Au, we find an overall high intensity in the two bands assigned with \ce{CH2} and \ce{CH3} bend vibrations in the alkyl chain, while in the bands which are characteristic for molecular orientation, we see a much more pronounced intensity contrast in the corresponding PiF channels. In these bands, high PiF intensities are seen on planar areas, while the valleys show much lower PiF (Figure~\ref{fig3}d). The pixel-wise correlations of PiF contrasts in the selected channels  confirm the different behavior of PiF intensities in the bands at 1581 and 1460~cm$^{-1}$ form those related to molecular orientation (Figure~\ref{fig3}e). Thresholds cannot be discriminated, which may result from the strong contributions of residual water vapor absorption in the spectral region of 1800 - 1350~cm$^{-1}$, which can be seen in the cluster mean spectra of the HCA analysis in Figure~\ref{fig3}f. The three clusters mean spectra for the clusters selected from the dendrogram (Figure~\ref{fig3}g) again vary by their intensities in bands characteristic for perylene core vibrations: HC2 shows the highest intensities in the two bands at $\nu=1324$~cm$^{-1}$ and $\nu=1291$~cm$^{-1}$, which are outside the spectral region influenced by water vapor absorption bands. The corresponding factor map (green) in Figure~\ref{fig3}h assigns this component to planar areas in the sample topography, while the cluster mean spectrum showing lowest intensities in the two bands (HC1) is assigned to the valleys (Figure~\ref{fig3}h, blue). Different to the HCA of PMIS-C8 on nanostructured Au (Figure~\ref{fig2}f-g), the cluster mean spectrum showing intermediate PiF intensities in the two bands (HC3) is not assigned to a particular sample structure; high factors of HC3 appear distributed over the whole sample area (Figure~\ref{fig3}h, purple). However, the factor maps of the two other clusters show some overlaps in the contour regions. It seems that three clusters were not enough to find the spectra related to the contours of the planar areas, or it is not possible to discriminate this area using an HCA, which may be a result of the strong water vapor lines in this data set. We had acquired two succeeding single frequency scans on the same sample area at $\nu=1650$~cm$^{-1}$ and $\nu=1694$~cm$^{-1}$ at two positions of PMIS-C8 on planar Au substrates. We used them to compare local band intensities by cropping the areas and presenting combined images using RGB colors; see Figure S6. However, the relative PiF intensity variation in the two sample areas does not show any distinct pattern. Thus, PMIS-C8 appears to align with its perylene core perpendicular to the planar sample surface, while vibrations parallel to the short and long axes of the perylene core cannot be discriminated in the monolayer films, in agreement with the observed dimer formation on Ag substrates.\cite{hupfer_supramolecular_2021} 

\begin{figure}[htb]
\centering
  \includegraphics[height=10 cm]{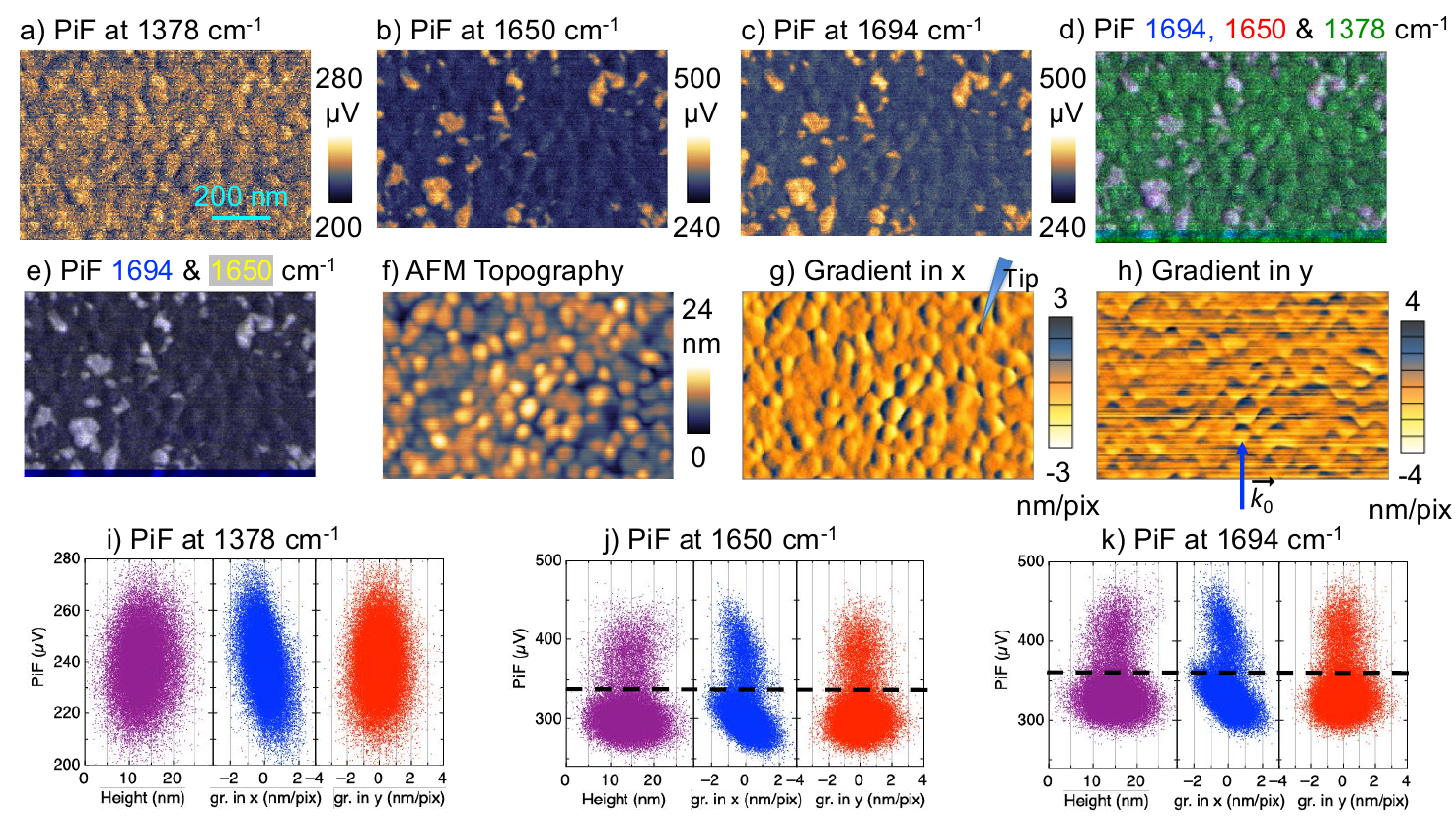}
  \caption{{\bf Molecular orientation of PMIS-C8 compared to field effects in PMIS-C8 monolayer on nanostructured Au:} a-c) PiF contrasts of a 1~$\mu$m wide area acquired at $\nu=1378$~cm$^{-1}$, $\nu=1650$~cm$^{-1}$ and $\nu=1694$~cm$^{-1}$, respectively; d,e) RGB images of cropped areas of PiF contrasts at d) 1694~cm$^{-1}$ (B), 1650~cm$^{-1}$ (R) and 1378~cm$^{-1}$ (G), and e) 1694~cm$^{-1}$ (B) and 1650~cm$^{-1}$ (R+G = yellow); f) AFM topography and g,h) gradients of topography calculated in g) horizontal ($x$) and h) vertical ($y$) direction. The blue shape in g) depicts the inclination of the scanning AFM tip. The blue arrow in h) marks the propagation $k_0$ of the incident field $E_0$. i-k) Correlation plots of PiF intensities in the three frequencies with sample topography and gradients, dashed lines in j,k mark the separation between subpopulations with high and low PiF intensities corresponding to $\mu \parallel E$ and $\mu \nparallel E$, respectively.}
  \label{fig4}
\end{figure}

\textbf{On nanostructured Au substrates, PMIS-C8 generally does not orient with its perylene core perpendicular to the Au surface,} as can be seen from scans of larger areas presented in Figure~\ref{fig4}. The PiF contrast acquired in a 1~$\mu$m wide area at $\nu=1378$~cm$^{-1}$ (Figure~\ref{fig4}a) reveals a complete coverage of the area by PMIS-C8 showing only moderate intensity variations related to the film surface structure (Figure~\ref{fig4}a). Subsequently acquired PiF contrasts  at $\nu=1650$~cm$^{-1}$ (Figure~\ref{fig4}b) and $\nu=1694$~cm$^{-1}$ (Figure~\ref{fig4}c) show varying intensities in these two absorption bands of the imide group with $\mu$ oriented parallel to the perylene core. A high PiF signal is visible only in several areas of this sample position extending over less than 100~nm. From this we conclude that the PMIS-C8 molecules generally orient with their perylene core parallel to the sample plane. The irregular surface structure and curvature of the evaporated Au substrate seem to prevent a large scale organization of PMIS-C8 in its preferred orientation perpendicular to the Au substrate, which has been found on planar Au (Figure~\ref{fig3}).

\subsection*{Field coupling effects}
To evaluate possible coupling effects of the film structure with the local photo-induced field, we derived gradients of the unprocessed sample topography (Figure~\ref{fig4}f) in horizontal ($x$) and vertical ($y$) directions; see Figure~\ref{fig4}g and h, respectively. Pixel-wise correlations of PiF intensities in the three frequencies with sample heights confirm the observation seen in the PiF contrasts of the scanned areas: for $\nu=1378$~cm$^{-1}$ a broad and unspecific distribution of PiF intensities is found. In the orientation-sensitive bands at 1650 and 1694~cm$^{-1}$, two subpopulations become visible: (i) large subpopulations showing low PiF intensities are seen below the dashed lines in Figures~\ref{fig4}j,k. They cover the full range of heights in the sample topography and correspond to positions with $\mu \nparallel E$; and (ii) smaller subpopulations showing higher PiF intensities appear above these dashed lines. The higher PiF intensities are mostly limited to height values in a narrower range around the mean heights (Figures~\ref{fig4}j,k). These subpopulations contain positions corresponding to $\mu \parallel E$.

In all three frequencies, PiF intensities increase with decreasing gradients in the horizontal direction ($x$). In the orientation-sensitive bands, the two subpopulations below and above the dashed lines in Figures~\ref{fig4}j,k increase separately. This general correlation between PiF intensities and gradients along $x$ is observed independent of the particular orientation of $\mu$ in respect to the orientation of $E$ for all three studied frequencies. It is caused by the hybrid field coupling with the probing AFM tip, as will be explained in the following: Negative gradients are assigned to the right hand slope of the "hill" structures. According to the manufacturer, the NCH-ppp tips used for scanning the samples, exhibit a tilt to the right, as depicted in Figure~\ref{fig4}g. The effect of this tip tilt on nanostructured surfaces was demonstrated by Anindo \textit{et al.}, who combined modeling and experiment to investigate the hybrid field coupling of the incident electric field $E_0$ with such a Pt-coated tip and a polymer nanosphere.\cite{anindo_photothermal_2025}
The observed correlation between PiF intensities and gradients for the PMIS-C8 monolayer on nanostructured Au is consistent with the reported increase of PiF intensities and modeled field $E$ at the right side of the nanosphere. An additional influence of the direction of the incident field $E_0$ from the bottom line of the scan (blue arrow in Figure~\ref{fig4}h) cannot be discerned. In the slow scanning direction, PiF intensities seem to be distributed symmetrically with positive and negative gradients. However, this influence might be concealed by noise in topography acquisition along the slow scanning direction, which is the cause of the horizontal lines seen in the gradient image in Figure~\ref{fig4}h.

\section*{Discussion}
From theoretical investigation combined with modeling it is known that the plasmonic/hybrid field enhancement between a metal tip and a metal/polymer film surface or nanostructure can reach several orders of magnitude, when polarized in the plane of incidence (p-polarization) and is much weaker for s-polarization.\cite{jahng_quantitative_2022, anindo_photothermal_2025, pascual_robledo_theoretical_2025} Theoretical considerations by Pascual Robledo \textit{et al.}\ show that in thin layers on highly reflecting surfaces the sensitivity to probe in-plane vibrations is reduced compared to probing out-of plane vibrations due to a vertical orientation of the near-field.\cite{pascual_robledo_theoretical_2025} Our PiF-IR investigation of oriented PMIS-C8 monolayer films on Au substrates provides an experimental evaluation of the modeled results by comparing PiF intensities in orientation sensitive bands related to vibrations of the perylene core\cite{del_cano_molecular_2004} and the imide group\cite{del_cano_molecular_2004} with PiF intensities of orientation-insensitive bands related to \ce{CH3} and \ce{CH2} bending\cite{coates_interpretation_2006} in the alkyl chain.

As expected, our PiF-IR results are able to reveal the orientation of PMIS-C8 with its perylene core perpendicular to the sample plane and the alkyl chain facing the air interface in LB monolayer films on planar Au substrate regions, and to some extent also on evaporated Au nanostructures. This observation agrees with a literature report on the macromolecular orientation of PMIS-C8 in LB films on plasmonically active Ag island substrates investigated using surface-enhanced resonance Raman spectroscopy (SERRS).\cite{hupfer_supramolecular_2021} Due to the high lateral resolution of PiF-IR ($\leq 5$~nm),\cite{shcherbakov_photo-induced_2025} we could additionally resolve the lateral structure of the PMIS-C8 films on both types of Au substrates on an almost single-molecular scale. From this, we found a reduced signal of characteristic bands related to vibrations of the perylene core in shallow $10-20$~nm wide valleys between planar Au regions, which were present on our planar Au substrates. In contrast, the PiF signal in bands related to the arbitrarily oriented alkyl chain showed only small intensity variations between the planar surface and valleys. On nanostructured Au substrates, we found variations in how PMIS-C8 was oriented on top of the Au islands formed during the evaporation of Au onto \ce{SiO2} substrates: On top of some of these "hills" a high PiF signal in the characteristic bands associated with imide group and core vibrations indicate an orientation of the molecules with their perylene core perpendicular to the substrate, similar to the observation on planar Au. However, regions showing this molecular orientation did not exceed over more than 3-4 neighboring islands. In most of the film, the characteristic bands associated with perylene core and imide group vibrations were absent or reduced. The cause of the varying molecular orientation on top of the nanostructured Au is not clear. A limiting factor on molecular alignment resulting from curvature of the Au surface may have some influence.

Our PiF-IR hyperspectral analysis of a region in the PMIS-C8 monolayer on nanostructured Au which showed high PiF intensities in the orientation-sensitive bands provides further insights into the local PMIS-C8 orientation in combination with field enhancement and orientation. For PiF channels related to isotropically oriented \ce{CH3} and \ce{CH2} bending of the alkyl chain ($\nu=1378$ and $1460$~cm$^{-1}$), we found enhanced PiF intensities on top of the "hills" in the nanostructure and a decrease with decreasing height of the topography. This observation agrees with the distribution of the photoinduced field intensities, which we obtained on a modeled nanostructure resembling one of the Au islands using finite element modeling (FEM). Applying a hierarchical cluster analysis, we found three distinctive subsets in the hyperscan: (i) in the top regions of the nanostructures, high PiF-intensities are seen in the orientation-sensitive bands. This agrees with the field enhancement observed also in the isotropic bands in combination with a molecular orientation of PMIS-C8 with its perylene core perpendicular to the Au surface; (ii) in the shallow regions of the sample, PiF spectra showed low absorption or absence of absorption in the orientation-sensitive bands while the intensity in the two prominent bands related to isotropically oriented alkyl chain vibrations ($\nu=1378$ and $1460$~cm$^{-1}$) was not altered, which clearly shows that the perylene cores were oriented rather parallel to the Au surface and not perpendicular to it in that region; and (iii) intermediate PiF intensities were found in the orientation-sensitive bands in the contours of the islands. According to the theoretical considerations by Pascual Robledo \textit{et al.}, also lateral field components are present at the apex of the metal tip.\cite{pascual_robledo_theoretical_2025} The field orientation on top of our modeled nanostructure is aligned perpendicular to the surface of the structure. Although our simplified model did not include any AFM tip, we assume that the coupled field between the tip and the nanostructure also contains a considerable component perpendicular to the surface of the nanostructure due to the lateral components present in the field close to the tip apex. Thus, we would expect to see also high PiF in the orientation-sensitive bands of PMIS-C8 in case the perylene core was oriented perpendicular to the substrate surface in this region. Since this is not the case, we conclude that the reduced PiF intensities are caused by molecular tilt in respect to the surface normal.

To study field coupling effects of the incident field with the nanostructure and between nanostructure and tip, we correlated PiF intensities obtained in three bands from PMIS-C8 on a $1\mu$m wide area with gradients of the local topography. PiF intensities in both anisotropically oriented imide vibration bands as well as at $\nu=1378$~cm$^{-1}$ (isotropically oriented \ce{CH3} and \ce{CH2} bending of the alkyl chain) showed an increase with increasing gradients from left to right in the horizontal fast scanning ($x$) direction. The scanning AFM tip is inclined by $\approx 30$\textdegree~from the right. We assign the observed asymmetry to an effect of field coupling between the inclined tip and the Au nanostructures enhancing the absorption in the thin PMIS-C8 layer on the nanostructure, which is consistent with an experimentally observed signal enhancement on a polymer nanosphere.\cite{anindo_photothermal_2025} In the vertical slow scanning ($y$) direction no asymmetry of PiF intensities with respect to the gradient could be discerned. However, the slow scanning direction is much more influenced by noise. We therefore suggest further experiments with varied fast scanning direction to investigate a mutual influence of field coupling also parallel to the field propagation vector $k_0$ of the incident field $E_0$.

\section*{Conclusions}
We used the high precision of PiF-IR to study local anisotropies in detected signals of an anisotropically absorbing LB monolayer on planar and nanostructured Au substrates. Our results  show a strong signal obtained from out-of plane vibrations on highly reflecting planar Au substrates, confirming theoretical considerations by Pascual Robledo \textit{et al.}.\cite{pascual_robledo_theoretical_2025} By comparing PiF intensities obtained from isotropically absorbing \ce{CH3} and \ce{CH2} bending vibrations of the alkyl chain with those related to orientation-sensitive molecular absorptions, we were able to discriminate effects caused by molecular orientation from those caused by coupling of the photo-induced field on nanostructured Au surfaces. By correlating PiF intensities in isotropically and anisotropically absorbing molecular vibration bands with gradients of the local topography we found an increase of intensities in the direction of the inclination angle of the scanning AFM tip in agreement with experimental observations of PiF intensities on a polymer sphere combined with modeling by Anindo \textit{et al.}.\cite{anindo_photothermal_2025}

Our results demonstrate the potential of PiF-IR to investigate local anisotropies on nanostructured surfaces and contribute to their understanding. The enhanced understanding will improve the interpretation of results obtained using nanoscale infrared spectroscopic imaging methods and enable their application to a wider area of nanostructured materials.

\section*{Data availability}
Data for this article,\cite{james_pif-ir_2025} including PiF-IR single frequency scans, spectra \& hyperspectra, complementary FTIR \& ATR spectra, AFM height \& phase data, a BAM series and interactive calculated IR spectra1 are available at  zenodo: 10.5281/zenodo.18060233.

The code for hyPIRana\cite{ali_hypirana_2026} can be found at https://doi.org/10.5281/zenodo.15270456 with DOI: 10.5281/zenodo.15270456. The code used for this study is also available as release v2.1.0 of our home-built software code hyPIRana on github: https://github.com/BioPOLIM/ hyPIRana.

\footnotetext{\textit{$^{*}$~E-mail: dantaube@gmx.de}}
\footnotetext{\textit{$^{a}$~Institute of Physical Chemistry (IPC), Friedrich Schiller University Jena, 07743 Jena, Germany.}}
\footnotetext{\textit{$^{b}$~Leibniz Institute of Photonic Technology (LIPHT), 07745 Jena, Germany.}}
\footnotetext{\textit{$^{c}$~Abbe School of Photonics, Friedrich Schiller University Jena, 07745 Jena, Germany}}
\footnotetext{\textit{$^{d}$~Institute of Organic and Macromolecular Chemistry (IOMC), Friedrich Schiller University Jena, 07743 Jena, Germany.}}
\footnotetext{\textit{$^{e}$~Jena Center of Soft Matter (JCSM), Friedrich Schiller University Jena, 07743 Jena, Germany.}}
\footnotetext{\textit{$^{f}$~Center for Energy and Environmental Chemistry Jena (CEEC Jena), Friedrich Schiller University Jena, 07743 Jena, Germany.}}
\footnotetext{\textit{$^{g}$~Institute of Condensed Matter Theory and Solid State Optics (IFTO), Friedrich Schiller University Jena, 07743 Jena, Germany.}}
\footnotetext{\textit{$^{h}$~University of Applied Sciences Landshut, 84036 Landshut, Germany.}}

\footnotetext{\dag~Supplementary Information available: [(i) Exemplary BAM images of a PMIS-C8 film formed at the water surface before film deposition, (ii) of pristine Au substrates and PMIS-C8 monolayer films, (iii) Topography images and PiF-contrasts of PiF-IR scan areas of the spectra presented in the main text together with two additional series of spectra on different sample positions, (iv) Further PiF contrasts of PMIS-C8 monolayers on planar Au, and (v) Further details on COMSOL modeling].}


\section*{Author Contributions}
Ayona James: Experimental, Data Analysis, Writing. 
Maryam Ali: Experimental, Data Analysis, Programming, Supervision. 
Zekai Ye: Experimental, Supervision, Writing. 
Phan Thi Yen Nhi: Computation, Data Analysis, Writing. 
Sharon Xavi: Modeling, Writing. 
Mashiat Huq: Modeling.
Sajib Barua: Experimental.
Yisak Tsegazab: Experimental.
Anna Elmanova: Computation.
Robin Schneider: Experimental, Data Analysis.
Olga Usitmenko: Experimental, Data analysis.
Sarmiza-Elena Stanca: Experimental.
Marco Diegel: Experimental.
Andrea Dellith: Experimental, Data Analysis. 
Uwe Hübner: Experimental, Methodology.
Christoph Krafft: Experimental.
Jasmin Finkelmeyer: Experimental, Methodology. 
Maximilian Hupfer: Experimental, Methodology.
Kalina Peneva: Conceptualization, Supervision, Project Administration, Funding Acquisition.
Matthias Zeisberger: Conceptualization, Modeling, Methodology.
Christin David: Modeling, Methodology, Supervision, Writing. 
Martin Presselt: Conceptualization, Methodology, Computation, Supervision, Writing. 
Daniela Täuber: Conceptualization, Writing, Methodology, Experimental, Data Analysis, Supervision, Project Administration, Funding Acquisition.

\section*{Conflicts of interest}
There are no conflicts to declare.

\section*{Acknowledgements}
For the VistaScope and the MirCat QCL financial support of the European Union via the Europäischer Fonds für Regionale Entwicklung (EFRE) and the Thüringer Ministerium für Wirtschaft, Wissenschaft und Digitale Gesellschaft (TMWWDG) is acknowledged (Projects: 2018 FGI 0023 and 2023 FGI 0018). DT acknowledges funding by a postdoctoral Scholarship (PolIRim) from Friedrich Schiller University Jena in 2020. DT and MA acknowledge funding from the German Research Foundation (DFG, Project: 542825796 "HiResi4RPE"). KP and OU acknowledge funding from DFG (CRC 1375 NOA, project number 398816777, Project B5). KP acknowledges funding from DFG in the framework of FOR5301 “FuncHeal” (project number 455748945, projects P3). The authors thank Beate Truckenbrodt and Ines Hirsch for providing access to the KSV NIMA Langmuir Blodgett trough in their education lab at the institute of Physical Chemistry, FSU Jena, when the otherwise used KSV 5000 was not available for LB film deposition.


\clearpage
\section*{SUPPORTING INFORMATION\\ Effect of hybrid field coupling in nanostructured surfaces on anisotropic signal detection in nanoscale infrared spectroscopic imaging methods}

\subsection*{Supporting Information abstract}
The Supporting Information provides: (i) Exemplary BAM images of a PMIS-C8 film formed at the water surface before film deposition, (ii) of pristine Au substrates and PMIS-C8 monolayer films, (iii) topography images and PiF-contrasts of PiF-IR scan areas of the spectra presented in the main text together with two additional series of spectra on different sample positions, (iv) further PiF contrasts of PMIS-C8 monolayers on planar Au, and (v) further details on COMSOL modeling.

\subsection*{Brewster angle microscopy (BAM) of a PMIS-C8 monolayer}
\begin{figure}[htb]
\centering
  \includegraphics[height=3.3cm]{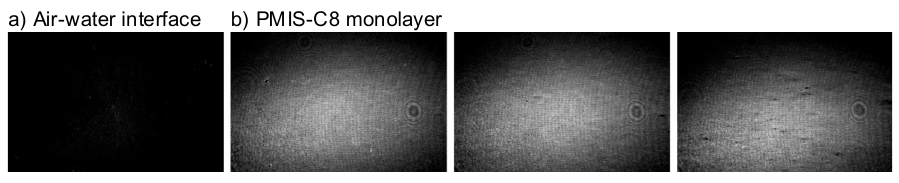}
  \caption{{\bf BAM images acquired during PMIS-C8 monolayer deposition:} a) air-water interface before adding PMIS-C8 solution and b) three different positions of a PMIS-C8 monolayer formed on the water surface at a surface pressure $\Pi=20$~mN/m.}
  \label{figS1}
\end{figure}
The cleanliness of the air-water interface before the addition of the PMIS-C8 solution is confirmed by the almost completely dark BAM image (Figure~\ref{figS1}a). At the final surface pressure of $\Pi=20$~mN/m, PMIS-C8 forms a homogeneous film on top of the water surface; see (Figure~\ref{figS1}b). Few areas show inhomogeneities, which appear as dark spots and lines, in particular, in the position shown in the rightmost image in Figure~\ref{figS1}b.

\subsection*{AFM results of pristine substrates and PMIS-C8 monolayer films}
\begin{figure}[htb]
\centering
  \includegraphics[height=7cm]{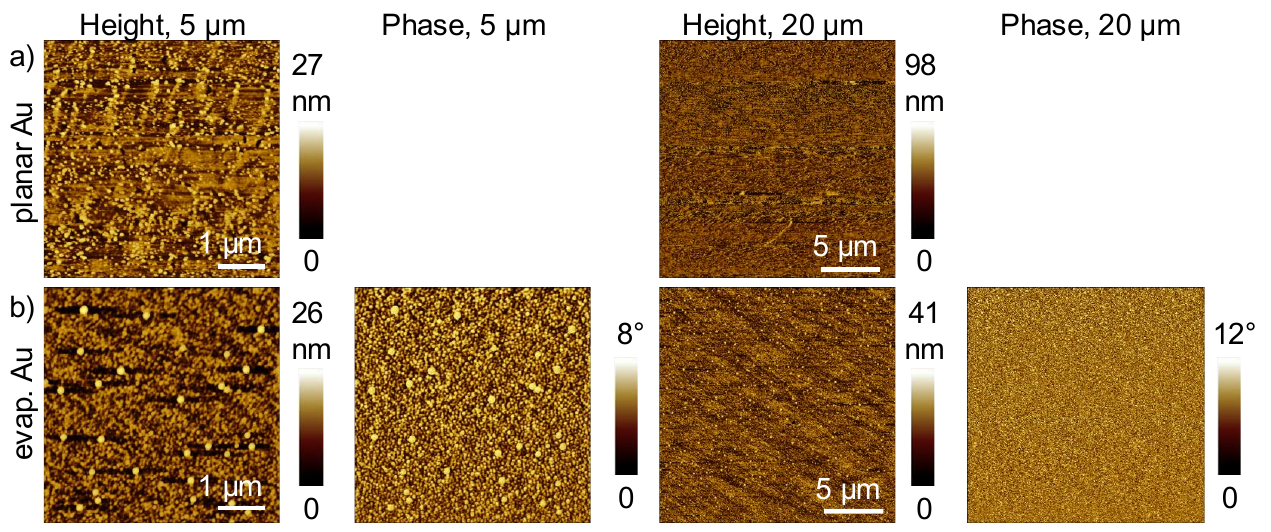}
  \caption{{\bf AFM results on pristine Au substrates.} 5~$\mu$m zoom (left) and 20~$\mu$m overview regions (right): a) height images acquired on planar Au and b) height and corresponding phase images acquired on evaporated Au.}
  \label{figS2}
\end{figure}

Unless they are kept sealed or under cleanroom conditions, pristine Au surfaces are highly attractive for any kind of charged dust particles. AFM height images of all Au substrates and film samples show small heights in the range of 10 - 20~nm apart from larger dust particles. The higher resolution zoom image acquired on pristine planar Au shows up to 1~$\mu$m wide planar areas which are surrounded  by small steps and some kind of debris (Figure~\ref{figS2}a, left). In contrast, the surface of the pristine evaporated Au substrate is covered by the typical island structure showing an irregular pattern of islands $\approx 15$~nm high and less than $\approx 100$~nm wide, as can be seen in the higher resolution zoom and phase images  in Figure~\ref{figS2}b, left.

\begin{figure}[htb]
\centering
  \includegraphics[height=14 cm]{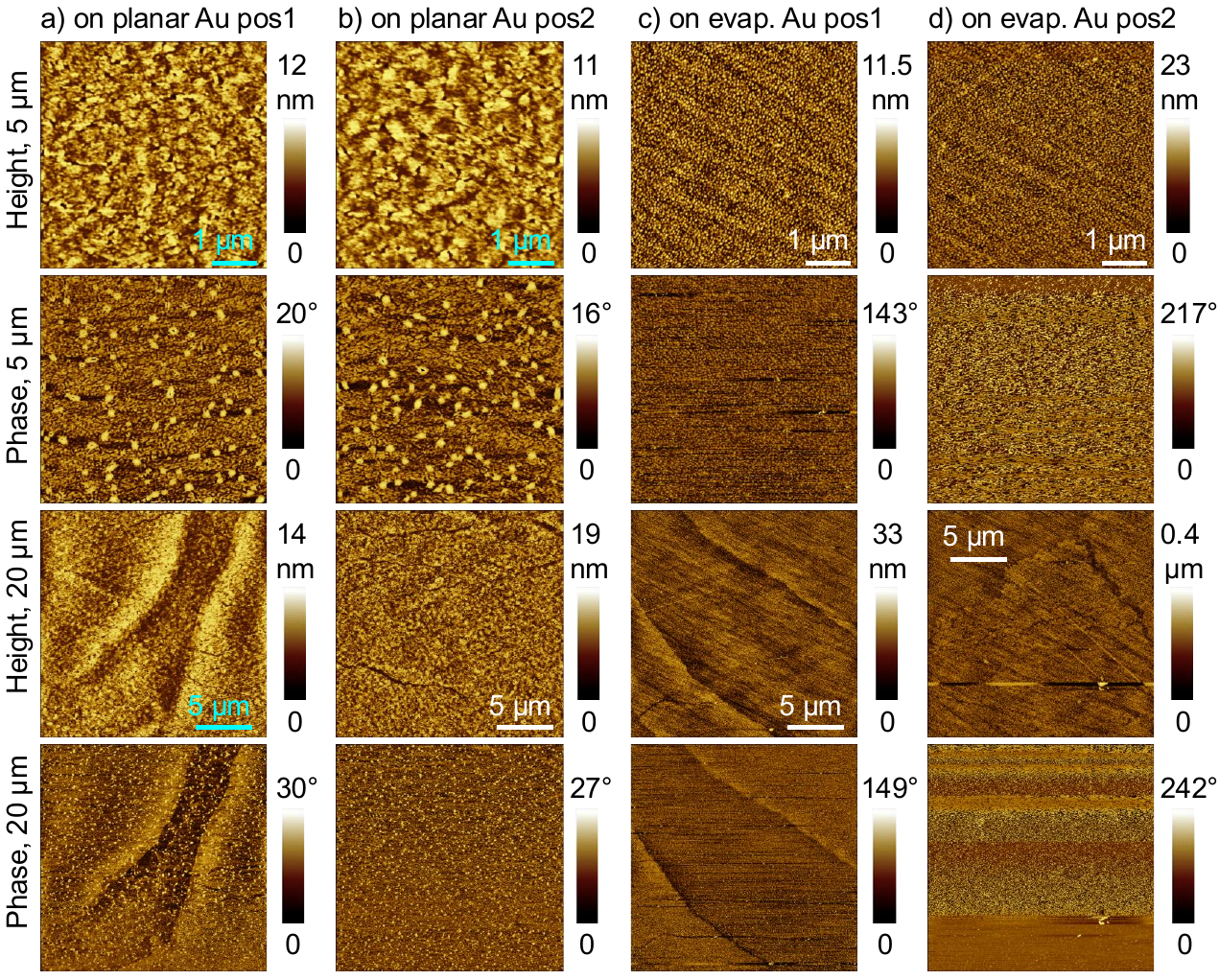}
  \caption{{\bf AFM results on PMIS-C8 monolayer films.}}
  \label{figS3}
\end{figure}
PMIS-C8 monolayers on planar Au show a pattern of up to several 100~nm wide planar flakes separated by small valleys and dips; see the higher resolution height images in Figures~\ref{figS3}a and b, top row. In the corresponding phase images, the deeper dips are enveloped by increased phase angles (Figures~\ref{figS3}a and b, upper middle row). The surface of the PMIS-C8 monolayers on nanostructured (evaporated) Au (Figure~\ref{figS3}c and d, top row) follows the island structure seen on the pristine evaporated Au substrate (Figure~\ref{figS2}b). However, the corresponding phase images of the two studied sample positions differ: apart from an $\approx 50$~nm wide area in the top region, the phase image of position 2 shows an irregular pattern, which is absent in the phase image of position 1; compare Figures~\ref{figS3}c and d, upper middle row. The  larger scale overview phase image of the second position shows horizontal stripes; see Figure~\ref{figS3}d, bottom row. In several cases, the pattern changes at positions showing larger dust particles. We therefore assume that the pattern in the phase images was caused by dust particles moved over the sample by the tip during scanning. The larger scale height images of both monolayer films in Figure~\ref{figS3} (lower middle row) show thin cracks with variable orientation and length, which can be assigned to cracks in the monolayer films in agreement with inhomogeneities seen also in the BAM images of the film surfaces before its transfer to the substrates (Figure~\ref{figS1}b).

\clearpage
\subsection*{Further details on PiF-IR spectra}
\begin{figure}[htb]
\centering
  \includegraphics[height=13 cm]{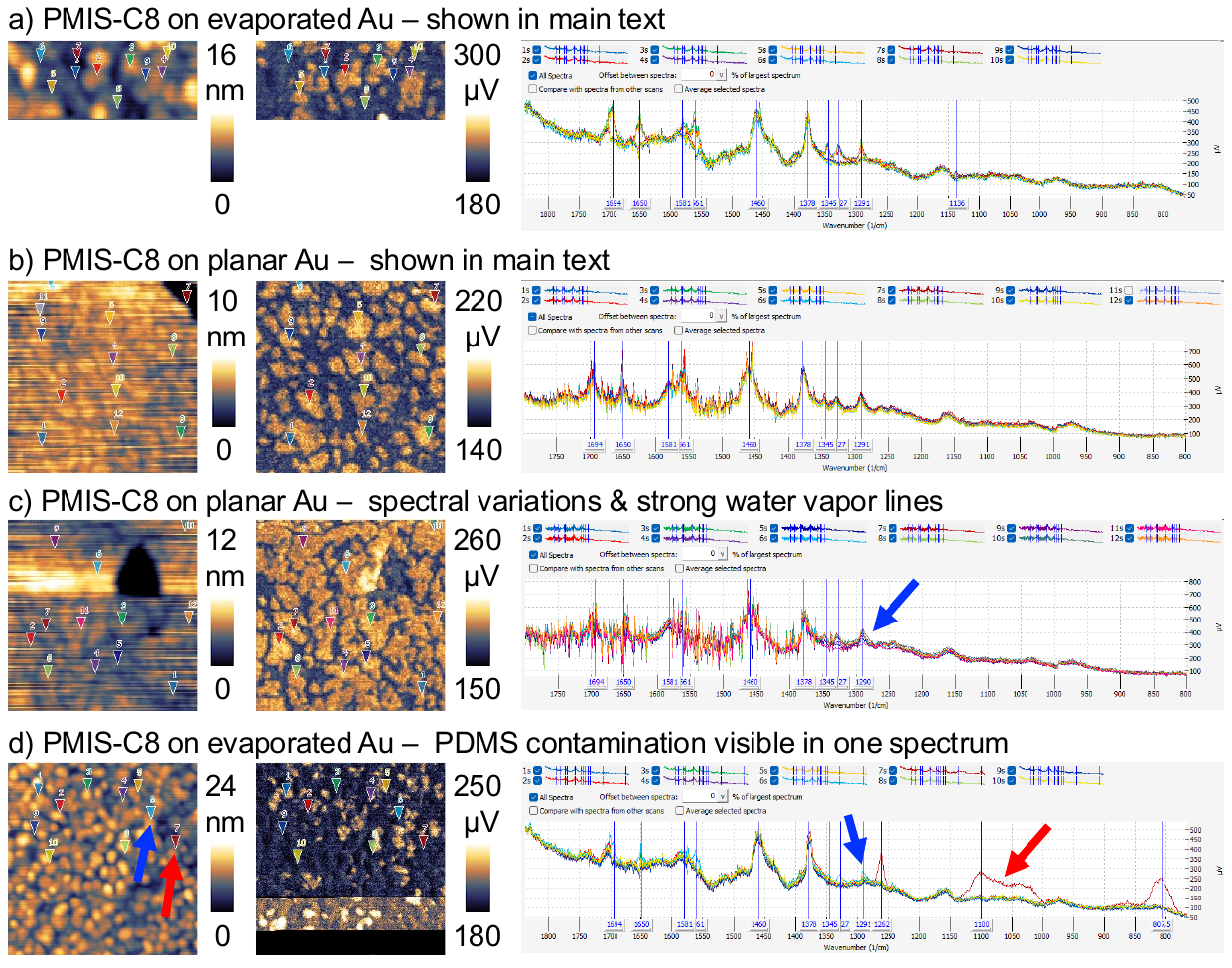}
  \caption{{\bf Positions of PiF-IR point spectra.} AFM topography images (left column) and PiF contrasts (second column) of scans at $\nu=1324$~cm$^{-1}$ with positions of point spectra marked in matching colors for PMIS-C8 monolayers on a) nanostructured Au, b) on planar Au, c) on planar Au in a different sample position to b; the blue arrow in the corresponding plot of spectra marks intensity variation in the band at 1291~cm$^{-1}$, and d) on nanostructured Au in a different sample position to a; the blue arrow marks the position with higher intensity in the characteristic core vibration bands and red arrows mark contamination with PDMS.}
  \label{figS4}
\end{figure}

PiF-IR spectra were acquired at several positions in previously scanned areas. Figure~\ref{figS4} shows AFM topography images (left column) and simultaneously acquired PiF contrasts at $\nu=1324$~cm$^{-1}$ (second left column) together with marked positions of point spectra. The right column shows the spectra in matching colors. The PiF contrasts of all sample positions show local variations in absorption at $\nu=1324$~cm$^{-1}$. The positions of the PMIS-C8 spectra on evaporated (nanostructured) Au discussed in the main text show a variety of low and high PiF intensities at 1324~cm$^{-1}$ (Figure~\ref{figS4}a). These spectra also show variations in the other absorption bands related to vibrations of the perylene core and imide group. In the two investigated sample positions of a PMIS-C8 monolayer on a planar Au substrate (Figures~\ref{figS4} b and c), positions showing high absorption at 1324~cm$^{-1}$ were selected for acquisition of point spectra. Consequently, the perylene core and imide group absorption bands are visible in these spectra. Few spectra in sample c show lower intensity in these bands. This data set suffers from strong residual water vapor However, the variation can be seen in the two lower frequency absorption bands at 1324 and 1291~cm$^{-1}$ (blue arrow).
The other data set acquired on the PMIS-C8 monolayer on an evaporated Au substrate mainly contains spectra acquired at positions showing low PiF intensities at 1324~cm$^{-1}$ (Figure~\ref{figS4}d). Only one spectrum shows higher intensities in the characteristic absorption bands related to perylene core and imide group vibrations (blue arrows). 

PiF-IR is highly sensitive to even small amounts of contamination.\cite{forster_quality_2025}. In our PiF-IR experiments, we frequently observed contributions of polydimethylsiloxane (PDMS), a highly volatile polymer that is used for storing AFM tips. PDMS could be easily discriminated in the data sets by its characteristic absorption bands at 1262~cm$^{-1}$, 1000-1150~cm$^{-1}$ and 807~cm$^{-1}$. An example is shown and marked with red arrows in Figure~\ref{figS4}d.

The bright and dark strips seen in the PiF contrast in (Figure~\ref{figS4}d) result from instabilities in the control of the illuminating MirCat QCL. The QCL turned off during the scan and was turned on again after we noticed the issue. However, its intensity changed to a higher value when it was turned on again, causing the brighter area. After a while, it turned off again and was not turned on until the scan was complete.

\subsection*{PiF contrasts of PMIS-C8 monolayers on planar Au}
\begin{figure}[htb]
\centering
  \includegraphics[height=9 cm]{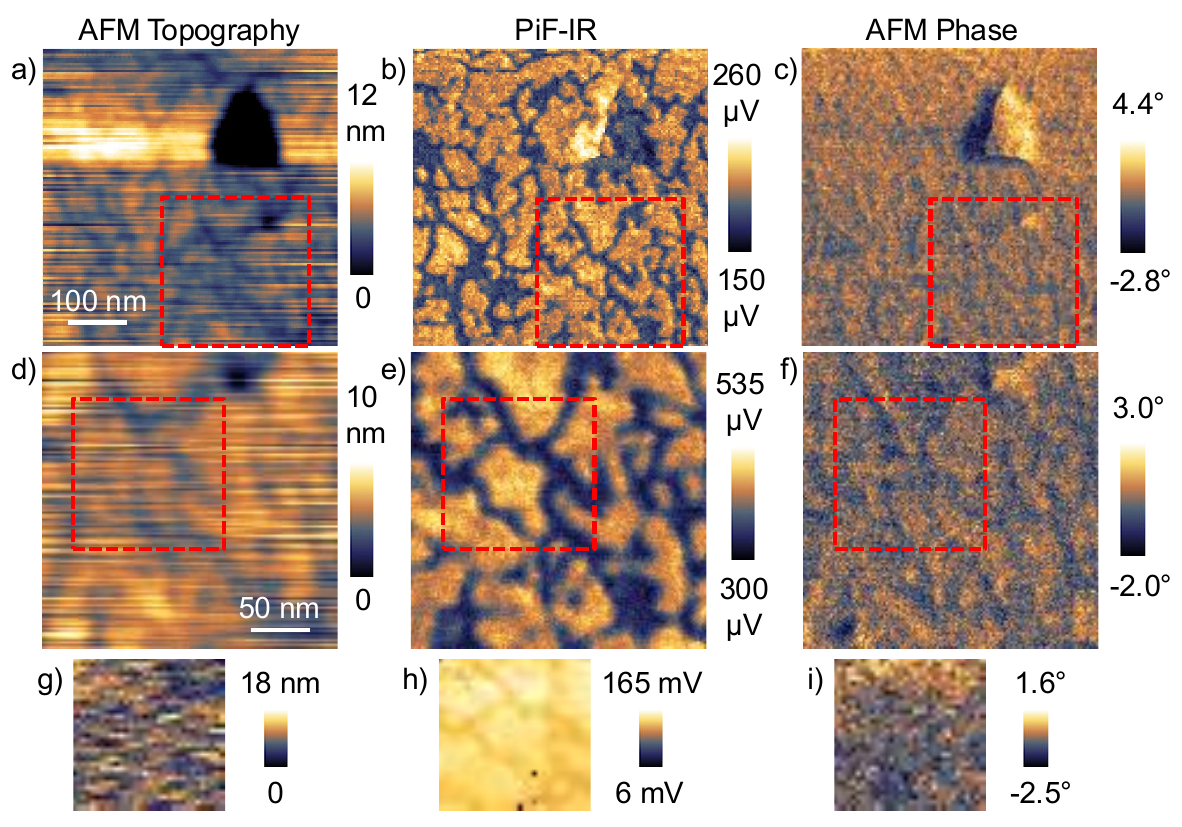}
  \caption{{\bf PiF contrasts of PMIS-C8 monolayers on planar Au.} a-c) overview scan, d-f) higher resolution zoom in area marked by red boxes in the overview scan images and g-i) hyperspectral scan in positions marked by red boxes in higher resolution scan: a, d, g) AFM Topography; b and e) single frequency PiF contrasts acquired at $\nu=1324$~cm$^{-1}$ and $\nu=1694$~cm$^{-1}$, respectively; c, f, i) simultaneously acquired phase images, and h) integrated PiF intensity over hyperscan spectral region.}
  \label{figS5}
\end{figure}

\begin{figure}[htb]
\centering
  \includegraphics[height=7 cm]{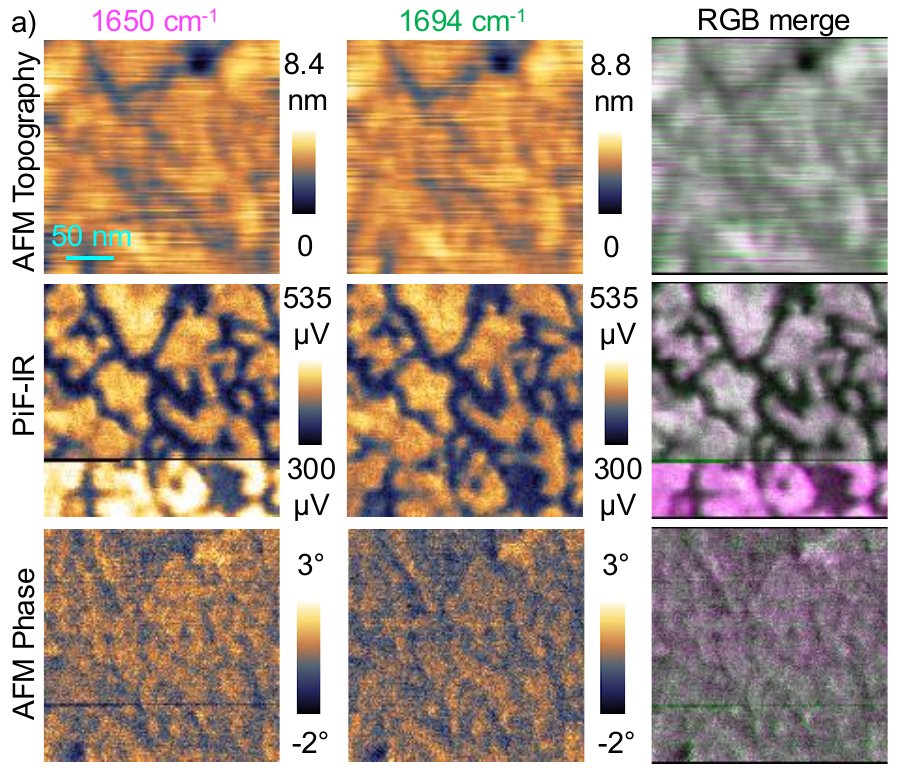}
  \includegraphics[height=7 cm]{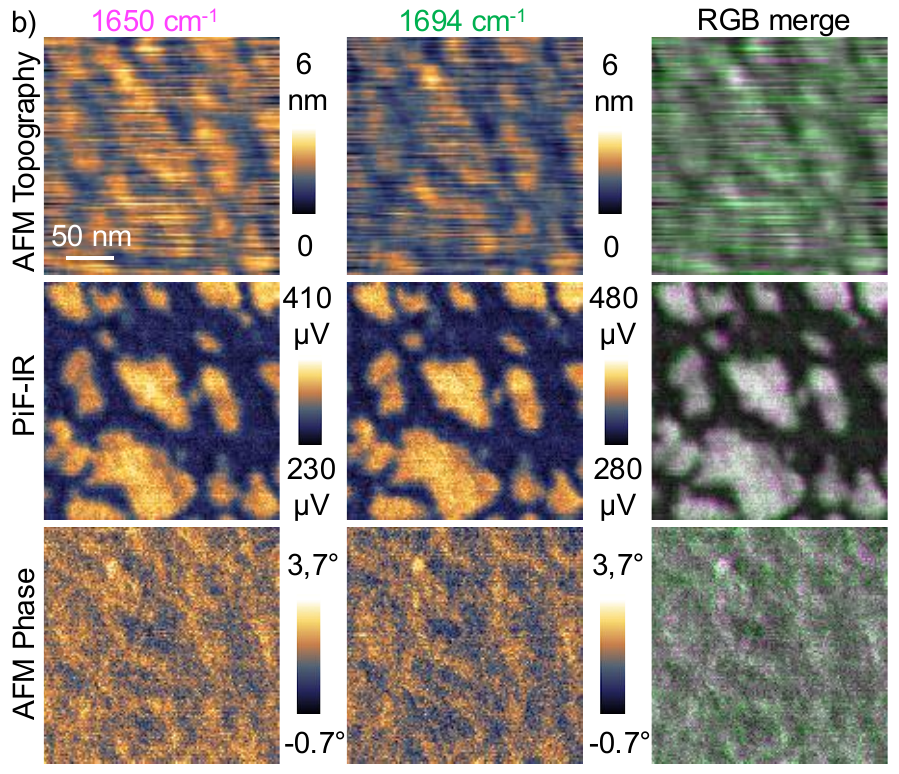}
  \caption{{\bf Comparison of PiF contrasts of PMIS-C8 monolayers on planar Au in the two bands related to symmetric and asymmetric \ce{C=O} stretch vibrations of the imide group.} AFM Topography images (top), PiF contrasts (middle) and simultaneously acquired phase images (bottom) acquired in two sample positions a) and b). Subsequent scans at $\nu=1650$~cm$^{-1}$ and $\nu=1694$~cm$^{-1}$ are presented as RGB merges with 1650~cm$^{-1}$ (R+B = pink) and 1694~cm$^{-1}$ (G).}
  \label{figS6}
\end{figure}
To study PMIS-C8 monolayer formation and molecular arrangement on planar Au substrates, we scanned a 500~nm wide area of the film surface. The AFM topography image shows an irregular pattern containing up to 100~nm wide planar structures surrounded by irregularly shaped valleys; see Figure~\ref{figS5}a. The scanned area also shows a deeper hole, in agreement with the larger area AFM images of PMIS-C8 monolayers on planar Au presented in Figure~\ref{figS3}a and b. The topography image was processed using line-wise image correction to compensate for tilt from sample mounting (the same scan is presented together with acquired sepctra in Figure~\ref{figS4}c). This resulted in an artifact showing bright stripes next to the hole. The PiF contrasts of the overview scan at $\nu=1324$~cm$^{-1}$ (Figure~\ref{figS5}b) and the higher resolution zoom at $\nu=1694$~cm$^{-1}$ (Figure~\ref{figS5}e) show higher values in the planar areas and lower signal in the valleys. This agrees with homogeneously oriented PMIS-C8 dimers in the planar areas. The integrated PiF contrast over the hyperscan (Figure~\ref{figS5}h) shows less contrast between planar areas and valleys, which can be explained by the small total contribution of the characteristic bands of perylene core and imide group to the total PiF signal over the spectral range of 1800 - 800~cm$^{-1}$ covered by the hyperspectral scan.

To compare contributions of asymmetric and symmetric \ce{C=O} stretch vibrations of the imide group, we acquired subsequent scans of the same sample position in two sample areas at the corresponding frequencies $\nu=1694$~cm$^{-1}$ and $\nu=1650$~cm$^{-1}$, respectively. The two scans were cropped to the same sample area and presented as RGB merge images (with 1650~cm$^{-1}$ (R+B = pink) and 1694~cm$^{-1}$ (G)); see Figure~\ref{figS6}. Due to the small hight difference and the homogeneous film material, the signal in the AFM height Figure~\ref{figS6} (top) and AFM phase images Figure~\ref{figS6} (bottom) is strongly influenced by instrument noise, resulting in horizontal stripes (fast scanning direction). During the scan at $\nu=1650$~cm$^{-1}$ on the first sample position Figure~\ref{figS6}a, the MirCat QCL turned off and was manually turned on again. This resulted in a black horizontal line followed by a higher intensity in the bottom area of the scan. All PiF contrasts show high intensity in the planar sample areas and lower intensity in the surrounding valleys, as can be seen in Figure~\ref{figS6} (middle row). The RGB merges of the two PiF contrasts show small intensity variations for the scans at the two different illumination frequencies. However, there is no distinct pattern. The variations are likely to be caused by instrument noise and a tiny offset between the pixel positions of the subsequent scans (scan resolution was 2 nm/pixel). The quite homogeneous PiF contrast in the two absorption bands agrees with the expected dimer formation of PMIS-C8 in monolayer films on metal substrates.\cite{hupfer_supramolecular_2021}

\subsection*{COMSOL modeling}
We modeled the optical response of two different gold surfaces: (i) planar Au and (ii) nanostructured Au under plane-wave illumination, using the \emph{Electromagnetic Waves, Frequency Domain} (EWFD) interface in COMSOL Multiphysics version 6.3 to study the plasmonic field of these surfaces for potential coupling with a metallic tip. All simulations were performed on a workstation with core processor Intel Core i7-14700KF CPU \@ 3.40 GHz with installed RAM of 64 GB.

\begin{figure}[htb]
\centering
  \includegraphics[height=7.5 cm]{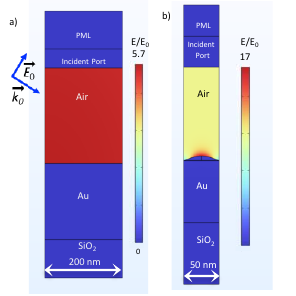}
  \caption{{\bf Electric field coupling in the air-gold interface:} a) modeled planar Au substrate and b) modeled nanostructured Au substrate with incident electric field $E_0$ at 60° in the $x$-$z$-plane.}
  \label{figS7}  
\end{figure}

Refractive indices for Au and \ce{SiO2} were taken from Olmon \textit{et al.},\cite{olmon_optical_2012} which is based on evaporated gold and measured in the region 0.300-24.93~$\mu$m, and Franta \textit{et al.},\cite{franta2016optical} which is valid in the spectral region 0.0248-125~$\mu$m, respectively. Both datasheets were imported from  refractiveindex.com~\cite{polyanskiy_refractiveindex_2025} and the COMSOL interpolation feature was used. The region between the incident port and the gold surface is assumed to be air. The field strengths obtained for some characteristic vibration frequencies of PMIS-C8 obtained in both geometries are presented in Table~\ref{Table2}. In the investigated spectral range of $\nu=1291$~cm$^{-1}$ to $\nu=1819$~cm$^{-1}$, the field strengths on both planar and nanostructured Au surfaces are almost constant with a very small increase with increasing frequency. The ratio of field strengths for both geometries remains constant at $1.6$.

\begin{table*}[h!]
\small
  \caption{Simulated normalized photoinduced fields on nanostructured and planar Au.}
  \label{Table2}
  \begin{tabular*}{1\textwidth}{@{\extracolsep{\fill}}cccccc}
    \hline\\
   $\nu$ (cm$^{-1}$) & $E_\text{nano-max}$  & $E_\text{nano-av}$ &$E_\text{planar-max}$ & $E_\text{planar-av}$
   & $E_\text{nano-av}/E_\text{planar-av}$ \\
    \\
    \hline\hline\\
   1291 &16.91
 & 5.73
& 6.95
& 3.55
 &1.61
\\
    \\
    1460 & 16.95
& 5.74
&6.97
& 3.57
 & 1.61
\\
    \\
    1557 & 16.93
& 5.76
& 6.97
& 3.58
 & 1.61
\\
    \\
    1581 & 16.93
& 5.75
& 6.98
& 3.58
 &1.61
\\
    \\
    1650 & 16.94
& 5.77
& 6.98
& 3.56
 & 1.62
\\
    \\
    1696 & 16.95
& 5.78
& 6.99
& 3.59
 & 1.61
\\
 \\
    1750 & 16.97
& 5.79
 & 6.99
 & 3.60
&1.61

    \\
    \\
    1819 & 16.98
& 5.80
 & 6.99
 & 3.61
&1.61
    \\
\\
    1850 & 16.99
& 5.80
 & 7.00
 & 3.61
&1.61
    \\
    
  \end{tabular*}
\end{table*}

The planar Au surface was modeled by placing a layer of 200~nm thick Au on 50~nm thick \ce{SiO2} and $L=200$~nm was used for the unit cell width; see Figure~\ref{figS7}a. The nanostructured Au was modeled using a $\cos^2(\pi*x/R)$ function with a radius of $R=25$~nm and a height of 15~nm and $L=2R=50$~nm was used for the unit cell width; see Figure~\ref{figS7}b. This 2D parametric curve was then revolved around the $z$-axis to obtain a 3D nanostructure. The nanostructure has a zero degree slope at its edge, which enables a smooth transition at the surface of the supporting 100~nm thick Au layer placed on a 100~nm thick \ce{SiO2} substrate. A tetrahedral volume mesh with surface refinement is used for the Au substrate and is applied along with an extreme fine mesh of 2.5~nm average size on the surface of the nanostructure.

We used Floquet periodic boundary conditions applied to all four surrounding surfaces to simulate a periodic array of these nanostructures. The periodic port is excited  with a TM polarized wave at an incidence of 60$^{\circ}$ with input port power of 1~W. PML layers are applied to the upper  boundary to ensure absorption of reflected waves at this interface. 

With these settings, we solve Maxwell's equations in the frequency domain using a fully coupled approach with the MUMPS direct solver at the frequency given via the wavelength through
$f = \dfrac{c}{\lambda}$. Convergence tolerances of the calculations are below the COMSOL defaults of $\leq 10^{-3}$\text{ relative}. The normalized values of the electric field for both planar and nanostructured Au are given in the table~\ref{Table2} showing that there is an increase in the field coupling with the nanostructured Au.

\bibliography{NanIR-anisotropy} 
\bibliographystyle{rsc} 

\end{document}